# Environmental performance of shared micromobility and personal alternatives using integrated modal LCA

Anne de Bortoli[a]*

[a] *University Paris-East, ENPC/Université Gustave Eiffel, Lab. City Mobility, Transportation, 6-8 Avenue Blaise Pascal, 77420 Champs-sur-Marne, France*

* *Corresponding author, anne.de-bortoli@enpc.fr*

## Abstract

The environmental performance of shared micromobility services compared to private alternatives has never been assessed using an integrated modal Life Cycle Assessment (LCA) relying on field data. Such an LCA is conducted on three shared micromobility services in Paris – bikes, second-generation e-scooters, and e-mopeds - and their private alternatives. Global warming potential, primary energy consumption, and the three endpoint damages are calculated. Sensitivity analyses on vehicle lifespan, shipping, servicing distance, and electricity mix are conducted. Electric micromobility ranks between active modes and personal ICE modes. Its impacts are globally driven by vehicle manufacturing. Ownership does not affect directly the environmental performance: the vehicle lifetime mileage does. Assessing the sole carbon footprint leads to biased environmental decision-making, as it is not correlated to the three damages: multicriteria LCA is mandatory to preserve the planet. Finally, a major change of paradigm is needed to eco-design modern transportation policies.

# 1. Introduction

## 1.1. Context

With food and buildings, transportation is one of the three key human activity sectors to environment preservation: it accounted for 14% of the global greenhouse gas (GHG) emissions in 2010 [1] on a scope 2 perimeter [2], i.e. excluding the impacts coming from vehicle manufacturing, maintenance and end-of-life as well as those from the infrastructure life cycle. In France, on the same perimeter, transportation accounted for 31% of the GHG emissions in 2019 [3], as well as 55% of the NOx emissions in 2018 [4]. As a strategic sector for planet preservation, transportation is the subject of numerous policy-driven and business-driven changes. Urban mobility systems have been strongly shaken these last fifteen years by new services, but also by specific crises.

Ten years ago in France, most Parisians moved on foot (60% of total trips) or used public transportation (27% of total trips), while cycling represented only 2% of the trips traveled [5]. Within a decade, shared mobility systems have emerged, such as the bike-sharing system Velib', running since 2007 [6], accounting in summer 2019 for around 12 000 bikes and 1300 stations [7]. While the cycling modal share expanded by 30% in the Parisian region between 2010 and 2018 [8], Velib' probably generated a drop in bike ownership: from 270 bikes per thousand of Parisians [5], the personal fleet dropped by 15% according to the first data of the 2018-2020 French mobility survey [8]. After a rapid expansion in the United States (US) [9], e-scooters (ES) have also boomed in Paris since June 2018: within one year, twelve free-floating electric scooter (FFES) operators emerged in the city, for a fleet volume estimated around 20 000 ES in the entire city in May 2019 [10], before the number of operators and vehicles being limited to respectively three and 15 000 by the City in 2020 [11]. Probably as a ripple effect, ES became the first micromobility market [12], the shared services impacting private acquisitions.

In 2020, 17% of the French big city's households are equipped with one or more personal standing scooters, two-thirds of these scooters being electric, and one-third being less than one year old [13]. The COVID-19 pandemic also made the FFES usage rise by 230% [14]. Shared e-mopeds appeared in Paris in 2015, but this new service usage rose by 27% during the pandemic [15], while the massive French public transportation strike in winter 2019-2020 also changed the modal share status quo, with 42% of the public transportation users switching for other modes: 17% for walking, 11% for the private car, 6% for car-sharing and 4% for cycling [14]. Vélib' recorded a 10% rise in memberships when the strike was announced [14]. Globally, shared mobility services tended to expand this last decade, despite a relative failure of free-floating bikes in Paris [16]. But while the sharing economy can be seen as a systematic way to promote sustainable societies [17], this belief can be easily challenged [18, 19].

### 1.2. Meta-objectives

The first objective of this study is to contribute to this belief challenge, specifically in the case of Paris but also in a worldwide context, by comparing the environmental performance of shared versus private micromobility modes. The second objective is to provide reference figures on the environmental performance of such modes including the impacts coming from the infrastructure and the complete vehicle life cycle, and to compare these modes with standard modes. The third objective is to appraise the contribution of the infrastructure and other components on the modal impacts, in order to give recommendations on the system boundary to be considered to design unbiased eco-friendly transportation policies; the system boundary being the selection of the components of a system to be considered in an LCA [20]. The fourth objective is to understand the implication for the planet preservation of supporting transportation policies through the sole spectrum of climate change and to rethink the "criteria scope" of mobility LCAs for environmental decision-making.

### 1.3. Paper's organization

After a literature review conducted on shared and/or micromobility environmental performance (section 2), the method and calculation process - including scenario analyses - to assess the environmental performance of the micromobility modes are presented (section 3). The data collection to develop Life Cycle Inventories (LCIs) in the case of Paris is detailed in section 4. Section 5 is dedicated to the environmental assessment results of shared and private micromobility modes in the case of Paris and the scenario analyses generalizing the results. Finally, section 6 discusses the carbon ranking of micromobilities compared with other modes in Paris and the US, before answering the question about the novel system boundary and criteria scope to choose to unbiasedly eco-design modern transportation policies.

# 2. Literature review

## 2.1. Cycling environmental performance

In the literature, a few studies address the environmental performance of micromobility and/or shared mobility. A comparative LCA was made on mechanical versus electric bicycles on a set of seven impact categories in a Swiss context with CML2001, CED, CExD, and IPCC 2007 characterization factors [21], but this approach did not consider the environmental impact from the infrastructure. Another study was conducted on bike-sharing systems and their climate change contribution in eight cities in the (US) [22]. The servicing distances lack reliability and regionalization, as they are based on an interview of one large US metropolitan area operator which is made anonymous and then extrapolated for the eight cities. Furthermore, the system boundary includes the shared stations but not the pavement. Bonilla-Alicea et al. [23] recently focused on the environmental impact of technological improvements on bike-sharing stations and shared bikes, looking at the three endpoint indicators (human health, resources, and ecosystems) as well as the climate change contribution, but again, the pavement impact is not accounted for. Dave conducted a hybrid LCA to assess few transportation modes in the US including mechanical and e-bikes but considering $CO_2$ exhaled by bikers, which is unusual, and mostly using environmental input-output LCA [24]. The environmental impact due to the infrastructure has been accounted for using the PaLATE tool, also based on a non-process based LCA and considering a generic pavement and not a cycle lane, as well as the allocation factors calculated by Chester for automobiles rather than bikes [25]. Her assessment leads to a carbon footprint of 33g of $CO_2$eq per passenger-kilometer traveled (pkt) both for cycling and walking, which is higher than other studies' assessments [19, 21].

## 2.2. E-scooter environmental performance

The impact of FFES on climate change – excluding the impact from the infrastructure - has been estimated by Hollingworth et al. in US conditions, based on local surveys and a material inventory of the microvehicle developed through the dismantling of a popular Chinese e-scooter model: the Xiaomi M365. The study includes an uncertainty analysis with a Monte Carlo simulation applied with assumed parameter distributions. Results show an average carbon footprint equal to 126g $CO_2$eq/pkt [26]. Dockless and private ES were also assessed for the city of Brussels on four midpoint indicators, using attributional LCA and the ReCiPe2016 characterization factors, reporting 131g of $CO_2$eq/pkt for the shared version and 67g of $CO_2$eq/pkt for the private version [27]. Even though the infrastructure is not integrated to the assessment of these two transportation modes, an integrated modal FFES LCA made in Paris reported smaller impacts – respectively 109 and around 60g of $CO_2$eq/pkt - with the same IPCC 2013 factors but using the consequential version of the ecoinvent database [19]. The previous FFES analyzed were first-generation ES, originally not designed for intensive use, and which will be called "entry-level ES". The impact of more robust ES, for private or shared used, has not been studied yet, while FFES operators now design their own stronger ES.

## 2.3. Motorcycle environmental performance

The assessment of shared motorcycles, either electric or with an internal combustion engine (ICE), are rare. Leuenberger and Frischknecht developed seminal LCIs for combustion and sitting electric scooters that are implemented in the ecoinvent database [28]. Interesting scenarios were developed to model the environmental performance of different kinds of motorcycles - including internal combustion and battery electric motorcycles – in 1990, 2017, and 2030, based on the World Harmonized Motorcycle Test Cycle model and European electricity mixes [29]. The smaller private e-motorcycle category can be used as a comparison to the shared e-mopeds in this Parisian study, on the full ReCiPe 2008 set of indicators. Electric mopeds and their energy consumption and GHG emissions were also assessed with a very fine dynamic consumption approach, restricted to the use stage of the mode [30]. Another publication restrained to the use stage estimates the energy consumption and carbon footprint of a 94-kg e-moped in Calcutta, India [31].

### 2.4. Synthesis

Finally, most of the shared and/or micromobility LCAs are incomplete: mostly in the life cycle stages and components of the transportation modes included in the system boundary, but also in the environmental issues addressed, often focused on climate change and potentially energy consumption. For instance, transportation LCA is known for a decade to need the inclusion of the infrastructure in the system boundary to perform unbiased analyses [32]. Nevertheless, excluding in one study [19], such an integrated approach has not been adopted yet for microvehicles. Moreover, field data from operators are still scarce, e.g. data on lifetime mileage, vehicle and infrastructure materials, consumption, and fleet management. With such data from the Paris' case, we propose to compare three shared micromobility modes and their private alternatives on the two most popular midpoint indicators as well as on the three endpoint indicators, before generalizing the shared micromobility results through extensive scenario analyses based on country-specific and non-country-specific parameters.

## 3. Method and calculation

### 3.1. Overview of the methodological choices

LCA is a method whose general framework and guidelines are established in ISO standards 14 040 and 14 044 [33, 34]. It is used to calculate the potential environmental impacts of a product, service, or system during its life cycle, related to a functional unit to allow system comparisons based on a similar unitary "quantity of function" provided. Initially developed and formalized for manufactured products, its application to transportation modes is more recent and has given rise to many studies and questions [25]. In this study, we perform a process-based LCA, using the OpenLCA open-source software and the consequential LCIs from the Ecoinvent V3.2 database. Ecoinvent is the most widespread and comprehensive background database to conduct LCA. It contains international industrial data quantifying the inputs and outputs flows (substances or energy) along the life cycle of industrial products or activities from the sectors of energy, manufacturing, transportation, chemistry, agriculture, etc.

### 3.2. Goal and scope definition

The "goal and scope definition" is the first step of any LCA following the ISO standards 14040 [35]. This LCA first aims to compare the environmental impact of three different shared micro-mobility services with their private alternatives in Paris: bikes, e-scooters, and mopeds. Then, it aims to generalize the environmental results of the three shared micromobility modes through sensitivity analyses. The functional unit in this study is "To carry one person over one kilometer", i.e. the pkt. We want to perform an integrated modal LCA approach, i.e. considering the environmental impacts coming from both the infrastructure and the vehicle life cycles: the system boundaries considered are illustrated in Figure 1. The life cycle of the vehicle includes manufacturing, use stage, servicing if any, maintenance, and end-of-life (EoL). The life cycle of the infrastructure includes the production, usage, maintenance, and EoL. Besides the specific infrastructure dedicated to bike-sharing systems that are described below, this modal LCA must consider the impacts of using the pavements and cycle lanes. The environmental performance of each mode will be calculated on Global Warming Potential (GWP), primary energy consumption, and damages on resources, human health, and ecosystems, respectively using the characterization methods IPCC 2013 GWP 100a (V1.03), Cumulative Energy Demand (V1.11), and ReCiPe Endpoint (E) (V1.10). The primary energy consumption is calculated by summing the different sources of energy consumption (fossil, nuclear, etc.). LCA indicator selection is always subjective. This set has been selected to cover all the environmental impact categories with a reduced number of indicators through the three aggregating endpoint indicators. We added the two most popular environmental indicators, e.g. GWP and energy consumption, to allow for literature comparisons. The calculation principle of these indicators based on ecoinvent data is described by Frischknecht et al. [36].

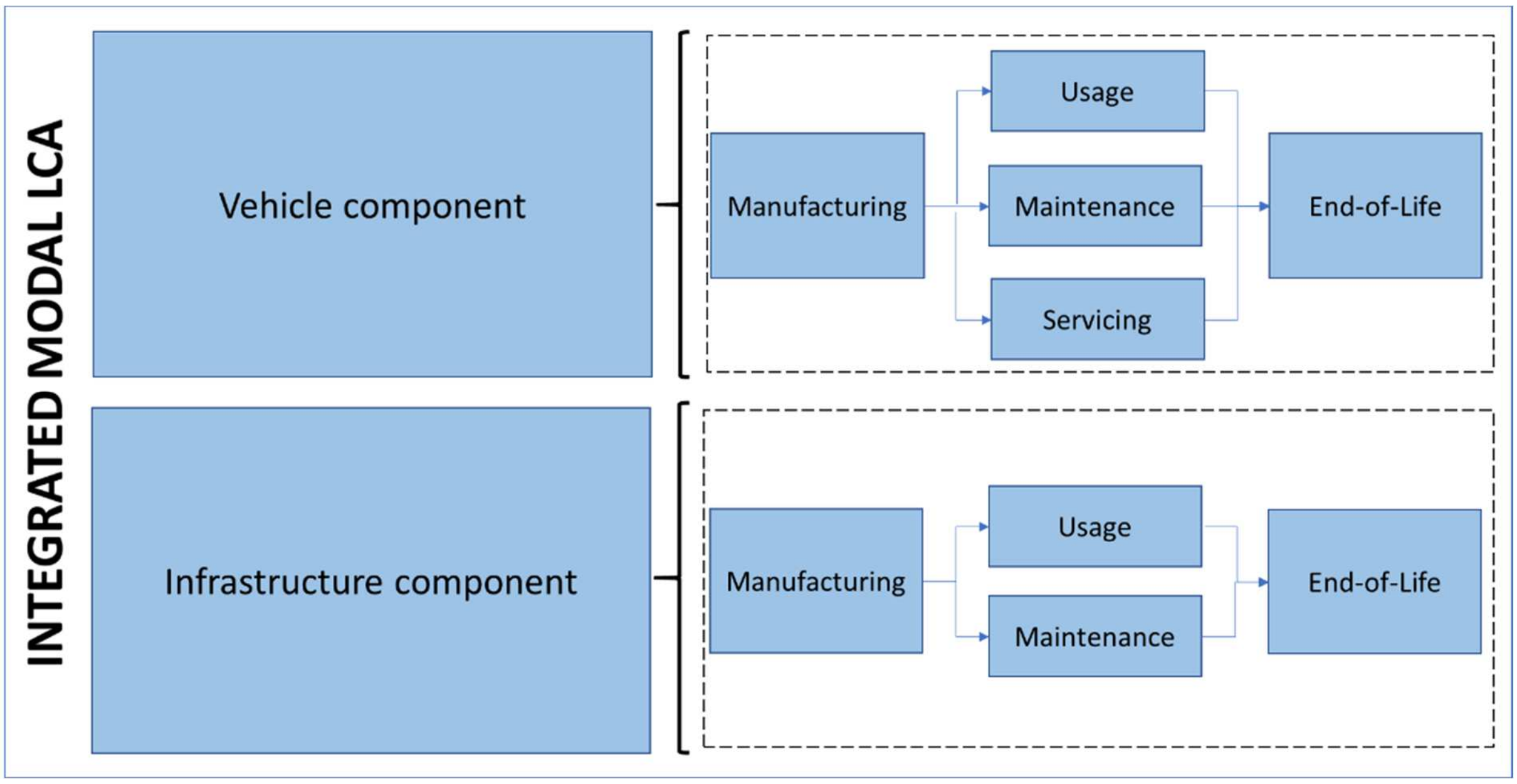

**Figure 1 Components and life cycle stages of the integrated modal LCA approach for transportation activities**

### 3.3. Calculation

**Integrated modal LCA**

Equation (1) calculates the environmental impact per pkt $\boldsymbol{EI_i}$ of each transportation mode *i,j* – using a vehicle type *i* and potentially different types of infrastructure, allocating the environmental impact of the infrastructure equally between the vehicles using it when the infrastructure is shared :

$$\boldsymbol{EI_i = EI_{veh\,i} + \sum_j EI_{i,infra\,j} = \frac{EI_{1veh\,i}}{PKT_{1veh\,i}} + EI_{usage\,veh\,1pkt\,i} + EI_{servicing\,veh\,1pkt\,i} + \sum_j \frac{1}{occupancy_i} \cdot \frac{1}{VKT_j} \cdot q_j \cdot EI_{1u\,j}} \quad (1)$$

with $\boldsymbol{EI_{veh\,i}}$ the environmental impact from the vehicle component of the mode *i* per pkt, $\boldsymbol{EI_{i,infra\,j}}$ the environmental impact from the infrastructure component *j* of the mode *i* per pkt, $\boldsymbol{EI_{1veh\,i}}$ the environmental impact for the manufacturing, maintenance and EoL of one average vehicle of the mode *i* over its life cycle, $\boldsymbol{PKT_{1veh\,i}}$ the number of pkt over the lifespan of the vehicle *i*, $\boldsymbol{EI_{usage\,veh\,1pkt\,i}}$ the environmental impact from the energy consumption of the vehicle for one pkt, $\boldsymbol{EI_{servicing\,veh\,1pkt\,i}}$ the environmental impact from the servicing of the vehicle for one pkt (only for shared vehicles), $\boldsymbol{occupancy_i}$ the average number of passenger per vehicle type *i*, $\boldsymbol{VKT_j}$ the annual number of vehicle-kilometers traveled (vkt) using the infrastructure *j*, $\boldsymbol{q_j}$ the number of units (surface, length, or item) of the infrastructure $j$ used by the mode $i$, $\boldsymbol{EI_{1u\,j}}$ the annual environmental impact of one unit of the infrastructure $j$.

In this study, we will consider the microvehicle occupancy equal to 1. The term $q_j.EI_{1u\,j}$ in Equation (1) represents the annual environmental impact of the entire type j infrastructural network used in the perimeter studied. In the case study, the perimeter is set for the city of Paris. The network size $q_j$ of each infrastructure type $j$ is indicated in the supplementary material.

## Infrastructure allocation factors

We recall the infrastructure allocation factors $\frac{1}{VKT_j}$ in the case of Paris' pavements and cycle lanes and their calculation parameters in Table 1. These infrastructure allocation factors mean that 4.39E-10 share of the pavement annual environmental impact is allocated to one vkt by motorcycle and 1.73E-9 share of the cycle lane environmental impact is attributed to one vtk by bike or ES. No allocation discrimination was made between vehicles based on mechanical damage or space consumption. Allocation discrimination is a methodological choice that is especially important in the case of low-trafficked pavements: in this case, the pavement environmental contribution can be substantial on the environmental impact of the mod, which is not the case in Paris [19].

Table 1 Annual road traffic and infrastructure allocation factor in Paris inner-city

<table>
<tr><th>Mode</th><th>VKT per vehicle type</th><th>Source</th><th>Infrastructure type</th><th>Infrastructure allocation factor $\frac{1}{VKT_j}$</th></tr>
<tr><td>Bus</td><td>4,71E+07</td><td>[37]</td><td rowspan="8">Road pavement</td><td rowspan="8">4,39E-10</td></tr>
<tr><td>Private car</td><td rowspan="2">1,85E+09</td><td rowspan="4">[5]</td></tr>
<tr><td>Taxi and ride-hailing</td></tr>
<tr><td>Shared motor scooter</td><td rowspan="2">3,74E+08</td></tr>
<tr><td>Private motor scooter</td></tr>
<tr><td>Truck</td><td>7,40E+05</td><td rowspan="2">[38]</td></tr>
<tr><td>Commercial vehicle</td><td>4,80E+06</td></tr>
<tr><td>Shared e-scooter</td><td>2,41E+08</td><td>[19]</td><td></td><td></td></tr>
</table>

| | | | | |
|---|---|---|---|---|
| Other two-wheeler (bike, personal ES) | 3,37E+08 | [5] | Cycle lane | 1,73E-09 |

**LCA computational structure**

The computational structure of LCA and especially the calculation of $\boldsymbol{EI}$ using the matrix model for inventory analysis is explained in detail by Heijungs et al. [39, 40], and accounts for loops in the consumption (e.g. the consumption of steel to produce steel). The life cycle inventory of a system lists the quantity of each input and output flow $\boldsymbol{flow_j}$ over the life cycle; these flows being product flows (e.g. the consumption of 1kg of steel), elementary flows (e.g. the emission of 2kg of fossil $CO_2$ in the atmosphere), or waste flows (e.g. the emission of 1m$^3$ of wastewater). The environmental impact $\boldsymbol{EI_k}$ of the system assessed on the impact category *k* is calculated, as presented in Equation 2, as the sum of the product of each flow $\boldsymbol{flow_j}$ by the characterization factor $\boldsymbol{CF_{j,k}}$ of the flow type j on the impact category *k*. This characterization factor represents the unitary environmental impact of $\boldsymbol{flow_j}$ on the impact category *k* based on a specific characterization method, such as ReCiPe.

$$EI_k = \sum_j flow_j . CF_{j,k} \quad (2)$$

## 3.4. Variability analyses

**Choice of parameters**

To assess the shared micromobility performance range, we will conduct scenario analyses. We choose to perform one-at-a-time sensitivity analyses rather than Monte Carlo simulations as some authors consider it is misleading to use it without perfectly known input distributions [41]. The vehicle lifespan and servicing were proved to be the key factors of the carbon performance of the shared first-generation ES in France and the US [19, 26]: we will conduct analyses on these non-country specific parameters. Moreover, the objective of the study is to estimate the environmental performance of micromobility from an international perspective. The shared micromobility market is expected to come mainly from the US, then from Europe and finally from China. The base-case scenario is focused on Paris, France. As vehicles - ES, bikes and motorcycles - are considered for the average global market, vehicle manufacturing impacts do not depend on the operating country. On the contrary, the electricity mix and vehicle shipping are two country-specific parameters to investigate.

**Vehicle lifespans**

The lifetime mileage is supposed to be reached when the vehicle breaks, or all the parts have been replaced. Four scenarios are proposed: worst-case, pessimistic, base-case, and optimistic. Figures are shown in Table 2. The base-case assumptions are explained in section 4. For the shared ES, the worst-case scenario is based on Quartz's day-lifespan estimate [42] multiplied by a factor 3 to reflect the new ES's stronger design, and average daily kilometers traveled in the study by Hollingsworth et al. [26]. For the other scenarios, as no data are available in the literature, we made assumptions based on users' forums, technical expertise, and operator declarations.

**Table 2 Lifetime mileage per vehicle type and scenario**

| Scenario | Worst-case | Pessimistic | Base-case | Optimistic |
|---|---|---|---|---|
| Shared bike | 2 000 | 7 000 | 12 500 | 20 000 |
| Shared e-scooter | 1000 | 2 000 | 7 300 | 13 650 |
| Shared e-moped | 8 000 | 20 000 | 48 000 | 65 000 |

**Servicing**

For the shared bikes, the optimistic scenario is set according to JC Decaux declaration, while the pessimistic and worst-case scenarios are taken from Luo et al. [22] for station-based bikes in the US. For the shared ES, servicing needs depend on the kind of batteries - swappable or not – and the warehouse location. In the case of Paris, warehouses are located outside the city and the batteries are not swappable: we consider the worst-case and pessimistic scenarios' assumptions based on a previous study of ES in Paris [19]. The base-case scenario is set according to the Parisian shared e-moped operator's declaration, as we are in the case of swappable batteries and a warehouse outside the city. The optimistic scenario estimates the reduction for swappable-battery vehicles and inner-city warehouse: instead of traveling from the suburbs to the city (average trip of 7 km), the average servicing can be cut by a factor two. This servicing distance will also be used for the optimistic scenario for shared e-mopeds, while the worst-case and pessimistic scenarios respectively consider twice and three times these distances.

| Scenario | worst-case | Pessimistic | Base-case | Optimistic |
|---|---|---|---|---|
| Shared bike (km/vkt) | 58 | 28 | 11 | 5 |
| Shared e-scooter (km/vkt) | 90 | 45 | 20 | 10 |
| Shared e-moped (km/vkt) | 60 | 40 | 20 | 10 |

**Electricity mixes**

We select regions where shared micromobility is used and which display a wide range of electricity environmental performances: average USA, China, Spain, the UK, Germany, Norway, the Netherlands, Australia, Denmark, Italy, and Canada.

**Shipping scenarios**

We want to test the environmental consequence of vehicle shipping, and propose the following scenarios according to options proposed by suppliers: sea + road, rail + road or air freight for Europe, sea + road, sea + rail or air freight for the US, and road or rail freight for China. Shipping distances are synthesized in Table 3. Air and sea shipping distances are estimated using the searates.com calculator. For sea shipping, the port of Shanghai, as the main port in China, is considered as the departure. The considered destination ports are Piraeus, Greece in Europe and Los Angeles, California in the US. Additional shipping distances by road in Europe and train in the US are respectively considered over 1500 km and 2500 km. Air shipping distances are estimated around 9000 km to Europe and 11 000 km to the US. The train + road shipping to Europe is modeled as follows: departure from Shenzhen by train (45% diesel trains and 55% electric trains) to Alanshankou in China (4 500 km), then diesel trains from Alanshankou to Troïskt, Russia (2 400 km), electric trains from Troïskt to Lodz, Poland (1 500 km), and 1000 km by truck. In China, shipping is considered on 1500 km, by train ((45% diesel and 55% electric) or truck.

**Table 3 Shipping scenarios by micromobility market**

| Market | Europe | | | USA | | | China | |
|---|---|---|---|---|---|---|---|---|
| Scenario code | UE 1 | UE2 | EU3 | US1 | US2 | US3 | CN1 | CN2 |
| Shipping scenario | sea + road | rail + road | air | sea + road | sea + rail | air | road | rail |
| Distances (km) | 14 250 + 1500 | 8 400 + 1 000 | 9000 | 11 120 + 2 500 | 11 120 + 2 500 | 11 000 | 1 500 | 1 500 |

# 4. Parisian data collection of Life Cycle Inventories

In this section, the lifetime mileages and LCIs for vehicle manufacturing, use stage, and (if applicable) servicing in Paris are described in detail. Maintenance and EoL stages are also considered but not specified: the EoL is included in manufacturing and the maintenance directly modeled using ecoinvent processes, as presented in the supplementary material. The calculation of the infrastructural impact $\boldsymbol{EI_{i,infra\,j}}$ uses LCIs that are extensively detailed in a previous study [19] and recalled in the supplementary material, excepting the impact from the bike-sharing system stations whose LCIs will be specified below in the bike LCIs section.

## 4.1. Bike LCIs

**Bike-sharing system**

The LCIs for the Velib' - the docked shared bikes in Paris - were collected and provided by the current vehicle and station supplier.

*Lifetime mileage*

The Velib' operator and supplier changed on January 1st, 2018, and data might not be consolidated. The annual mileage is estimated at around 10 000 km for each bike. 19 000 bikes are currently operated, while 40 000 bikes have been manufactured so far. We can then estimate an average lifespan of 31 months/40 000*19 000 = 14.7 months, thus a lifetime mileage of 10 000/12*14.7 = 12 250 km. This result must be underestimated due to bikes stolen, and their consequent second lives.

*Manufacturing and end-of-life*

The mechanical bikes weigh 20.6 kg, while the e-bikes weigh 27 kg. Only the electric equipment differs between the two versions, with a 2.4-kg lithium-ion battery and a 4-kg motor. In summer 2020, the fleet consists of 40% e-bike and 60% mechanical bikes, but around 50% of the kilometers would be traveled with the electric version according to the fleet manufacturer. The lithium-ion battery is under warranty for a 70% residual capacity after 500 charging cycles, representing 14 000 km traveled at a speed of 20 km/h. Thus, no battery replacement is needed over the lifetime mileage.

*Use stage*

The energy consumption depends on the trip topography and will be estimated at around 1 kWh/100 km [28]. Batteries are charged at low voltage.

*Servicing*

The rebalancing distances were not provided by the Parisian operator, but are estimated based on data from a company operating in other French cities. Rebalancing relies on a 1180-kg tractor electric vehicle with a 380-kg trailer vehicle and a 376-kg lithium-ion battery. The distance traveled is estimated at around eleven meters per pkt: this operator owns 28 vehicles rebalancing 11 000 bikes by traveling each day 65 km, and its bikes travel 5 500 km a year.

*Stations*

According to the bike provider, the 1400 Velib' stations represent a total surface of 92 000 m² in Paris. LCIs of the sidewalk life cycle come from a previous study [19] and are recalled in the supplementary material. Each station is equipped with : (a) one e-kiosk (for customers), whose LCIs are adapted and completed from Bonilla-Alicea et al. [23] and detailed in the supplementary material, and (b) a total of 46 500 docks made of 23 kg of steel and 0.5kg of plastic each. The station lifespan is considered equal to 10 years, i.e. the duration of the operator's contract. There are different station technologies. According to the Parisian operator in 2005, a standard 20-dock station consumed 3651 kWh a year in Paris. Current stations in Paris counting around 46500/1400~33 docks, we will consider a prorate of 6063 kWh/station.year, thus an annual consumption of 6064*1400=8.49 MkWh/year. As a sensitivity analysis, we will consider the following alternative: according to the former Parisian operator, a 20-dock low-consumption station consumes as low as 879 kWh a year due to technological enhancement, thus a alternative total Parisian consumption of 2.04 MkWh/year. These stations are annually used by 19 000 bikes traveling 10 000 km a year, thus an allocation factor of 1/(19000*10000)=5.29E-09 of the annual station impact.

**Personal bikes**

*Lifetime mileage*

The personal bike lifetime mileage in France is estimated around 20 000 km, based on responses to a survey dedicated to the use of bikes and e-scooters and conducted in big French cities in 2020 [13] (see calculation in the supplementary material). This lifespan is 50% higher than the estimate for the shared bikes.

*Manufacturing*

The personal bikes will be considered as a 17-kg mechanical bike, despite approximately 3% of the Paris region personal fleet were e-bikes in May 2020 [43], and possibly more today due to the effect of the Covid-19 pandemic on the Parisian mobility system [13]. The bike has an aluminum frame and additional equipment such as carriers and lights. It is manufactured in China and transported to Europe [28].

### 4.2. E-scooter LCIs

Two e-scooters are considered: an entry-level model (first generation ES) and a mid-range model (second-generation ES). The personal ES can be either of them, while the shared version is a mid-range model.

**Shared e-scooter**

*Lifetime mileage*

The shared e-scooters are supposed to last 24 months, as declared by most of the operators equipped with second-generation e-scooters [44, 45], with a distance traveled a day equal to (100-66%) x 18 miles [26], i.e. 10 km, thus a 7 300 km lifetime mileage.

*Manufacturing*

LCIs for the shared ES are adapted from a previous study [19]. For the production stage, from the first generation of e-scooters weighing 12 kg each, the quantities are estimated to represent the second-generation of shared e-scooters, weighing approximately 22 kg with a double-capacity battery, the same equipment and electronic components, and a rescaled frame weight and assembly consumptions calculated with a ratio between the remaining weight after removing similar components and the battery [44]. New LCIs are indicated in the supplementary material.

*Use stage*

The e-scooter consumption is considered equal to 0.335 kWh/20 km, according to a previous study [19]: we can consider that doubling the e-scooter weight do not change notably the consumption as it would represent (22-12) / (12+70) = 12% additional weight with one average user weighing 70kg.

*Servicing*

We select the scenario considering Light Commercial Vehicles (LCV) traveling over 90 km with 100 ES aboard from a previous study [19] but switching from an ICE vehicle to the same electric LCV used for the bike-sharing system. With the double capacity range of the batteries, we consider that the ES is picked up after having traveled 20 kilometers instead of 10, the servicing distance per pkt being 90 km/100ES/20km, e.g. 45 m/pkt.

**Personal e-scooter**

*Lifetime mileage*

The lifetime mileage of the entry-level model is supposed to be 4 000 km based on web users' declarations. The personal mid-range e-scooter is assumed to last 10 000 km according to user feedback.

*Manufacturing*

Manufacturing LCIs of the entry-level model come from the same previous study [19]. The mid-range model manufacturing LCIs are the same as those of the shared ES.

*Use stage*

The consumption of the personal ES is supposed to equal the shared ES consumption.

### 4.3. Moped LCIs

**Shared e-mopeds**

The data to model shared e-mopeds come from exchanges carried out in 2020 with the major company operating in Paris. This company owned 3750 e-mopeds in Paris in mid-2020, operating for almost 5 years.

*Lifetime mileage*

The operator expects an 8-year lifespan, with an annual mileage of 6000 km, i.e. 48 000 km over the moped life cycle.

*Manufacturing*

The ecoinvent processes selected and their related quantities are indicated in the supplementary material. The mass of the e-moped is 98 kg, plus 28 kg of lithium-ion batteries (one fixed and one swappable batteries), for a total capacity of 4 kWh. Over the 48000-km e-moped lifespan, the battery must be changed once.

*Use stage*

The consumption is 3.3 kWh/100km, and the swappable batteries are charged with low-voltage electricity, in dedicated warehouses in the Parisian suburb.

*Servicing*

Servicing is done using electric LCV, weighing 1250 kg + 350 kg of lithium-battery, with an estimated lifespan of 150 000 km, and consuming 25 kWh/100km. The battery is changed once over the LCV lifespan. 25 LCV are in operation each day, traveling 50 km/day. Each kilometer traveled by an e-moped is then supported by a 20-m servicing, to maintain the whole moped and charge its swappable battery.

**Personal motorcycles**

*Lifetime mileage*

The average personal Parisian motorcycle is considered to have a 50000-km lifetime mileage [28].

*Manufacturing*

The average personal motorcycle is an ICE motorcycle. A Parisian mobility survey showed that, in 2010, it was a combination of 65% of mopeds, 29% of heavier motorcycles, and 6% of other motorized two-wheelers [46]. We will consider a mix of 2/3 of a moped weighing 90 kg and 1/3 of a heavier motorcycle weighing 200 kg, for an average motorcycle weighing 127 kg.

*Use stage*

The use stage is modeled using the HBEFA 4.1 calculator [47], for an average motorcycle in France in 2020. The simulated baseline consumption is equal to 39g of petrol per kilometer, i.e. 5.4 L/100km (gasoline density = 0.72) which seems to be overestimated for our average motorcycle. Indeed, a moped consumes around 2.3 L/100 km and a motorcycle 6.1 L/100km in urban areas in France [48]: a ratio of (2/3*2.3+1/3*6.1)/5.4=0.66 will be applied to correct linearly both the consumption and emissions. Fuel consumption and European regulated tailpipe emissions (HC, NOx, PM, CO, and $CO_2$) per kilometer are indicated in the supplementary material. Other emissions are considered through the fuel production modeled with the ecoinvent process.

### 4.4. Synthesis of the vehicle model main parameters

The main characteristics of the parameters used in the vehicle models are presented in Table 4. It highlights that all the shared vehicles studied present shorter lifespans than their private alternatives: -4% for mopeds, -27% for the same model of ES, and -38% for bikes. The reliability of these estimates could nevertheless be improved by specific surveys. The e-scooter lifespans are based on operator and user declarations, and especially less reliable (or more variable) due to the novelty of emerging technologies and services and consequential lack of data. The shared bikes and e-moped lifespans are also based on operator self-declarations. In the case of the e-moped, it is a simple estimate, as the service only runs in Paris for five years. In the case of the shared bikes, another operator estimates that the 10 000 km traveled a year might have been slightly overestimated. The shorter servicing distance for the free-floating e-moped than for the shared ES is explained by the swappable batteries on the mopeds.

**Table 4 Synthesis of the main parameters of the vehicle modeling**

| Vehicle | Ownership | Weight (kg) | Lifetime mileage (km) | Servicing vehicle | Servicing distance (m/pkt) | LCI source |
|---|---|---|---|---|---|---|
| **Bike** | Shared | 21/27 | 12 500 | Electric LCV | 11 | Author |
| | Personal | 17 | 20 000 | N/A | | Ecoinvent |
| **e-scooter** | Shared | 22 | 7 300 | Electric LCV | 45 | Author |
| | Personal – entry-level model | 12 | 4 000 | N/A | | Author |
| | Personal – mid-range model | 22 | 10 000 | N/A | | |
| **Moped** | Shared | 136 | 48 000 | Electric LCV | 20 | Author |
| | Personal | 127 | 50 000 | N/A | | Author |

N/A=not applicable

# 5. Results

## 5.1. Environmental comparison between micromobility modes

Figure 2 is a spider chart that facilitates the multicriteria performance comparison of the different micromobility modes. It uses the results from Table 5, mathematically normalized based on the most impacting mode on each environmental indicator. It shows that the most efficient mode depends on the indicator considered: the average motorcycle is on average twice as much impacting as the other modes on all indicators except for damage to human health, where shared e-mopeds and e-scooters are more impacting (around 2.3-2.5.10-6 DALY/pkt). This is due to (a) the high contribution of the lithium-ion battery manufacturing to this damage category, and (b) the rather high ratio between the battery weight and the lifetime mileage for these modes compared to the other electric modes (between 2.32 10-4 and 1.17 10-3 kg/pkt, against 9.6 10-5 kg/pkt for the shared bikes). The personal motorcycle does not include any lithium-ion battery, thus ranks fourth most efficient mode on this damage category, with the same impact as the mid-range private ES. The private bike is, by far, the most efficient mode on all indicators, followed by the shared bike on the three damage types, but ranked equal with the shared e-scooter in terms of carbon footprint and also equally-ranked with the mid-range private ES in terms of primary energy consumption. The primary energy consumption of all modes except the private motorcycle and private bike is included in a narrow range – between 0.938 and 1.31 MJeq/pkt. Entry-level personal ES and shared ES show very similar impacts on the five indicators. Except for the human health damage, the shared e-moped presents very good environmental performances considering its potential speed (50 km/h), comfort, and safety, especially compared to ES and motorcycles. It is ranked second on the primary energy consumption performance and carbon footprint after the private bike, third on the ecosystem damage after shared and private bikes, and fourth on resource damage after the bikes and the mid-range private ES.

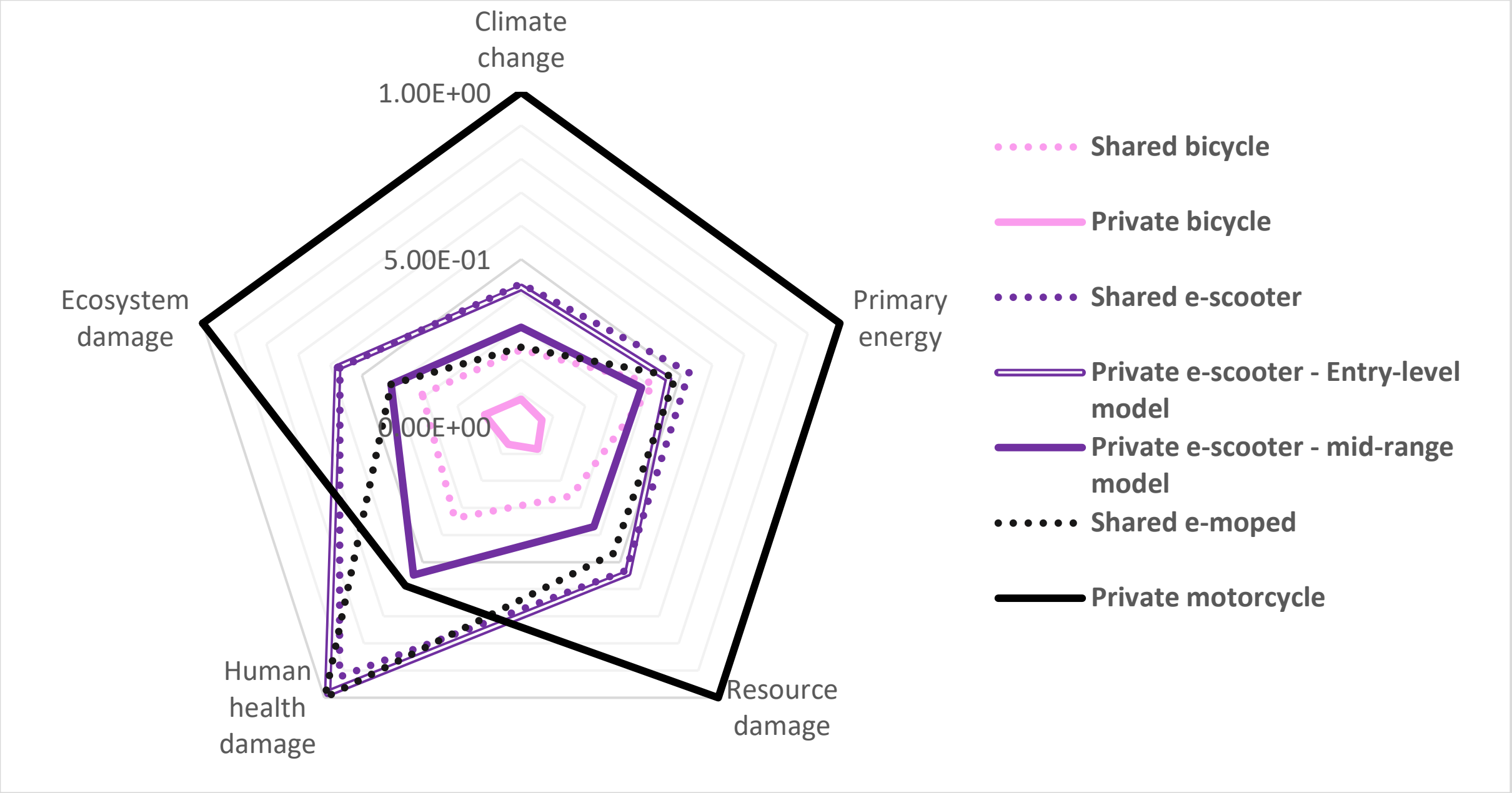

**Figure 2 Comparison of normalized environmental impacts of the Parisian micromobility modes**

**Table 5 Environmental impacts of the different shared and private micromobility modes in Paris**

| Mode | Climate change ($kgCO_2eq/pkt$) | Primary energy (MJeq/pkt) | Resource damage ($/pkt) | Human health damage (DALY/pkt) | Ecosystem damage (species.year/pkt) |
|---|---|---|---|---|---|
| Shared bicycle | 3.29E-02 | 1.04E+00 | 2.41E-03 | 8.31E-07 | 1.25E-09 |
| Private bicycle | 1.17E-02 | 1.59E-01 | 7.72E-04 | 1.61E-07 | 4.65E-10 |
| Shared e-scooter | 6.10E-02 | 1.31E+00 | 5.01E-03 | 2.28E-06 | 2.29E-09 |
| Private e-scooter - Entry-level model | 5.95E-02 | 1.15E+00 | 5.10E-03 | 2.43E-06 | 2.33E-09 |
| Private e-scooter - mid-range model | 4.24E-02 | 9.38E-01 | 3.48E-03 | 1.35E-06 | 1.65E-09 |
| Shared e-moped | 3.40E-02 | 1.20E+00 | 4.42E-03 | 2.47E-06 | 1.64E-09 |
| Private motorcycle | 1.43E-01 | 2.49E+00 | 9.44E-03 | 1.45E-06 | 4.03E-09 |

The GHG savings brought by the new station technology described in the bike LCIs and that diminishes the electricity consumption saves 1g of $CO_2eq/pkt$, i.e. a 3% drop in the carbon footprint of the mode. Moreover, in the baseline Parisian study, the average bike frame is considered made of aluminum, a material that is more harmful to the environment than steel. A steel bike emits 35kg of $CO_2$ over its life cycle [49], while the personal aluminum bike in this study emits 212kg of $CO_2eq$. Ceteris paribus, personal bikes made of steel can cut the carbon footprint of the mode by a factor three, to 3.5g of $CO_2eq/km$.

### 5.2. Life cycle stage and component contributions

The contributions of each life cycle stage and component to the different impacts are presented in this subsection. Detailed figures focusing on GWP, primary energy consumption (EC), and the three areas of environmental protection - damage to resources (DR), human health (DHH), and ecosystems (DE) - can be found in the supplementary material. The contributions and influent flows are analyzed below, before discussing an extract of the results presented in Figure 3.

**Bikes**

For the shared bike mode, vehicle manufacturing accounts for 70% of the GWP. Surprisingly, the e-bike is only 24% more emitting than the mechanical version. The Velib stations account for 24% of the carbon footprint due to the electricity consumption (42%), the e-kiosk (32%) and the dock (26%) manufacturing, the cycle lane contribution being around 3.5%. The vehicle use stage is also negligible on the carbon footprint, accounting for 0.3% of the total life cycle emissions, while the servicing accounts for 2%. On EC, the infrastructure is the most impacting component, with 68% of the total impact, 95% due to the stations, and especially the direct electricity consumption (98%). 27% of the consumption is due to the vehicle life cycle, and only 2 and 3% are due to the use and servicing stages. The contribution to DE is also mostly carried by the vehicle (65%) and the Velib' stations (25%, half of that due to the electricity consumption), while the cycle lanes account for 7%, the servicing for 2.4%, and the electricity consumed by the bikes 0.4%. The contributions are quite similar on DHH, with 67% of the impact from the vehicle and 23% from the stations, but the servicing is more impacting (8%) and the cycle lanes contribution lower (2%). The electricity consumed by the bikes still brings a negligible contribution (0.5%). Finally, on DR, the vehicle manufacturing and station contributions are also major (65 and 27%) while the cycle lanes, stations, and use stage only account for respectively 5, 4, and 0.3%. The bike manufacturing and the station consumption are thus the two key parameters of the bike-sharing system environmental impacts.

In the case of the private bike, the vehicle is still the most impact component: respectively 93%, 82%, 85%, 93%, and 89% of GWP, EC, DE, DHH, and DR. The cycle lanes account for the rest. Bike manufacturing carries most of the impacts, the maintenance contribution being below 10%. Let's notice the weight of the vehicle carbon footprint: 10.5g of $CO_2$eq/pkt for a 17-kg bike lasting 20 000 km. The life cycle of a 27-kg e-bike emits around 320kg of $CO_2$eq on its life cycle. It equals to 26g of $CO_2$eq/pkt for a 12500-km lifespan and 16g of $CO_2$eq for a 20000-km lifespan. Most of the impact (76%) comes from the production of the mechanical bike, the electric motor accounting for 18% of this impact, and the battery for only 3%. The maintenance accounts for 8% of the emissions.

**E-scooters**

For the shared second-generation ES, most of the impacts come from the vehicle life cycle too: for example, it represents 79% of the GWP, 50% being due to the aluminum alloy. The servicing, using e-vans and optimized routes, accounts for 9% of the carbon emissions on the life cycle. The pavement life cycle brings 10% of the GHG emissions. Electricity consumption contributes negligibly to climate change (2%). On the energy aspect, the vehicle life cycle accounts for 43% of the consumptions, while respectively 18% and 17% are due to the ES consumption and the servicing. The pavement brings 23% of the impact, due to the bitumen which represents 18% of the total life cycle impact. In the three areas of protection, the vehicle life cycle is also the key parameter, accounting respectively for 67%, 72%, and 64% of the DE, DHH, and DR. The aluminum alloy explains this phenomenon on DE and DR, but the printed wiring board and the battery cell production are more important (resp. 24 and 22%) on DHH, the aluminum accounting for 19% of the total DHH of the mode. The pavement accounts for 20 and 22% of the total damages to DE and DR, due to the bitumen for DR, and to the aggregates for DE. On DHH, it contributes to 4% of the modal impact, while the servicing takes 21% of the total, mainly due to the e-van lithium-ion battery (15%).

For the personal entry-level ES mode, the vehicle life cycle accounts for 78% of DE, mainly due to the aluminum alloy production (41%). The pavement brings 20% of the impacts, and the ES electricity consumption is still marginal (2.5%). The contributions to DR are almost the same, but 24% of the impact comes from the vehicle is explained by the aluminum, the rest mainly being due to the electronics (18% for the transistor, 12% for the battery, 8% for the charger, 8% for the printed wiring board, 7% for the electric motor). The vehicle contribution to DHH is even more important (94%), and mainly due to the electronic components (41% due to the printed wiring board, 18% to the battery). In terms of energy, the vehicle, infrastructure, and use stage accounts for respectively 54%, 26%, and 21% of the total EC. The carbon footprint is driven by the vehicle component (88%) and especially the aluminum alloy (50%). The battery contribution is marginal (3.5%). The pavement life cycle accounts for 10% of the total GHG emissions, and the use stage for 2%.

The personal mid-range ES mode is less impacting than the entry-level ES mode. The vehicle component brings respectively 83%, 44%, 68%, 90%, and 67% of GWP, EC, DE, DHH, and DR. Globally, the same sub-contribution explanations as for the other ES modes apply, but the vehicle contribution decreases as the lifetime mileage increases, while the use stage and infrastructure component take mechanically a higher share of the impact (25 and 31% on the primary energy indicator).

**Motorcycles**

The shared e-moped mode presents contributions rather different from the other modes: on EC, only 28% of the impact comes from the vehicle, while the use stage represents 39%, the servicing 9%, and the pavement 25% of the total impact. On the GWP, the vehicle component still brings a large part of the impact (68%), the infrastructure accounting for 18%, the servicing and the use stage for 7% each. On the vehicle component (excluding the use and servicing stages), 55% of the impacts are due to the scooter manufacturing plus 36% to the battery, 23% to the maintenance stage, the recycling of the battery canceling 15% of the vehicle life cycle. 85% of the servicing carbon footprint comes from the e-van production, with almost equal contributions from the van manufacturing, battery production, and van maintenance. Damage mainly comes from the vehicle (69 to 86% depending on the category). DHH is mainly due to battery manufacturing (73%), but also to the scooter manufacturing (28%). The toxicity midpoint impact accounts for 96% of the endpoint damage. The moped maintenance only accounts for 1% of the total damage. The pavement is the second most important contributor for DE (28%) and DR (25%). Servicing is second on DHH, with 9% of the impact.
Just as standard ICE modes, the main contributor of the personal ICE motorcycle mode impact is the use stage (80% on GWP, 74% on EC, between 61 and 70% for damages). 23% of the modal carbon footprint comes from the upstream GHG emissions to produce gasoline. 16% of the carbon footprint is explained by motorcycle manufacturing (11%) and maintenance (5%). The pavement only brings 4% of the total GHG emissions. We find the same order of magnitude for the three damage categories.

**Synthesis**

Globally, aluminum brings a high contribution to GWP, DE, and DR of micromobility modes. Electronics - and especially batteries and printed wiring board – are harmful to human health. Vehicle servicing with e-vans keeps the impact from this stage rather low in France. The pavement presents substantial contributions, around 10% to GWP and 20-30% to EC, DE, and DR depending on the mode. Bitumen is the major contributor to GWP and DR, while aggregates strongly contribute to DE. The use stage of e-micromobility shows a negligible contribution to the environmental impacts in France.
Figure 3 shows on three very different micromobility modes – the shared ES, the shared e-moped, and the private gasoline motorcycle – that the contribution pattern to the total impact of a mode depends on the type of mobility assessed and the environmental indicator considered. For the shared ES, contribution patterns are quite similar between GWP and DHH, while the contribution patterns are quite different for the e-moped and private motorcycle on these two indicators. The DR and DE contribution patterns are almost identical for the motorcycle mode, themselves not very different from the EC pattern. The same type of similarity is found between the DR and DE contribution patterns for the e-moped, although very different from the pattern for the motorcycle. This analysis highlights the importance of multicriteria integrated modal LCA to properly capture the environmental key performance factors of transportation modes, and design policies considering burden shifting.

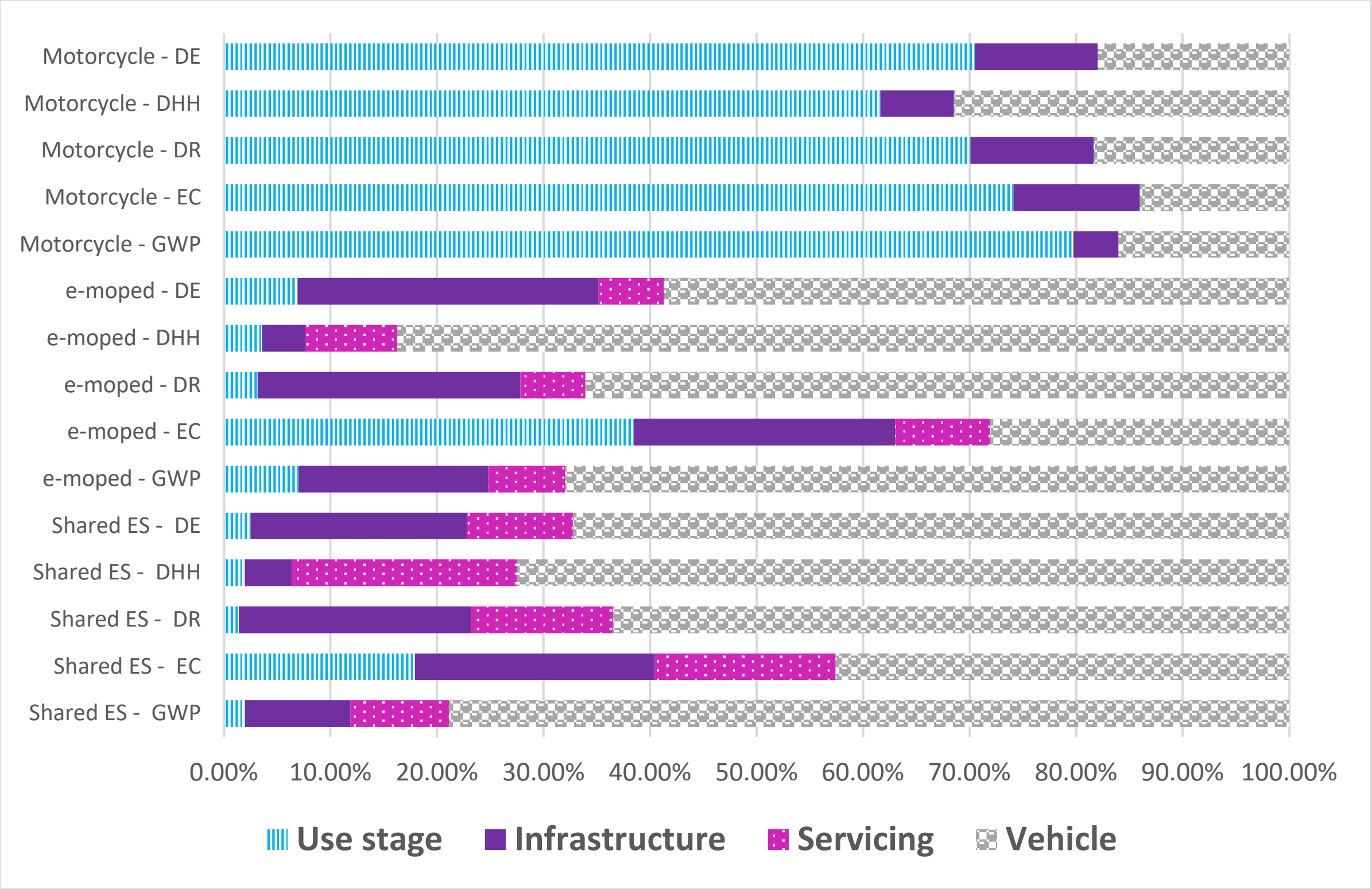

**Figure 3 Selected contribution analysis of the vehicle, infrastructure, and use stage to the five environmental indicators on the total impact of three modes: shared ES, shared e-moped, and private motorcycle**

## 5.3. Sensitivity analyses

### Vehicle lifespan

Figure 4 presents the carbon footprint of the three shared modes depending on the lifespan scenarios defined in section 2. Similar figures for the four other indicators are presented in the supplementary material. These results highlight how environmental ranking relies on lifespan. For instance, the shared e-scooter is less emitting in the optimistic scenario than the shared bike or e-moped in the pessimistic scenario. Moreover, except for the shared e-scooter in the worst-case scenario, the carbon footprint of shared micromobility is always lower than the one of car mobility (around 200g of $CO_2$eq/pkt) [19]. More specifically, additional calculations set the minimal lifetime mileage to emit less than the average car: 1500 km for the shared bike, 1800 km for the shared ES, and 6000 km for the shared e-moped. Nevertheless, expanding the lifespan reaches a limit to lower the modal carbon footprint: an unrealistic 200 000-km lifespan leads to respectively 11g, 15g, and 16g for the shared bike, ES, and e-moped. Due to the impacts from the other life cycle stages than vehicle manufacturing, only improving the lifespan will not make micromobility less emitting than e-public transportation in Paris, accounting for less than 10g of $C0_2eq$/pkt [19]. On other indicators, the ranking is quite similar, excepting for EC. On this indicator, the shared bike's base-case scenario presents similar performance to the optimistic scenarios for shared ES and e-mopeds. It can be explained by the fact that, with the shared bike, more than 50% of the kilometers are fueled by food calories, an energy that is not accounting for in this LCA.

Finally, the shared bikes and e-mopeds perform generally better than the shared ES, while these latter perform generally a bit better than the e-mopeds within one type of scenario.

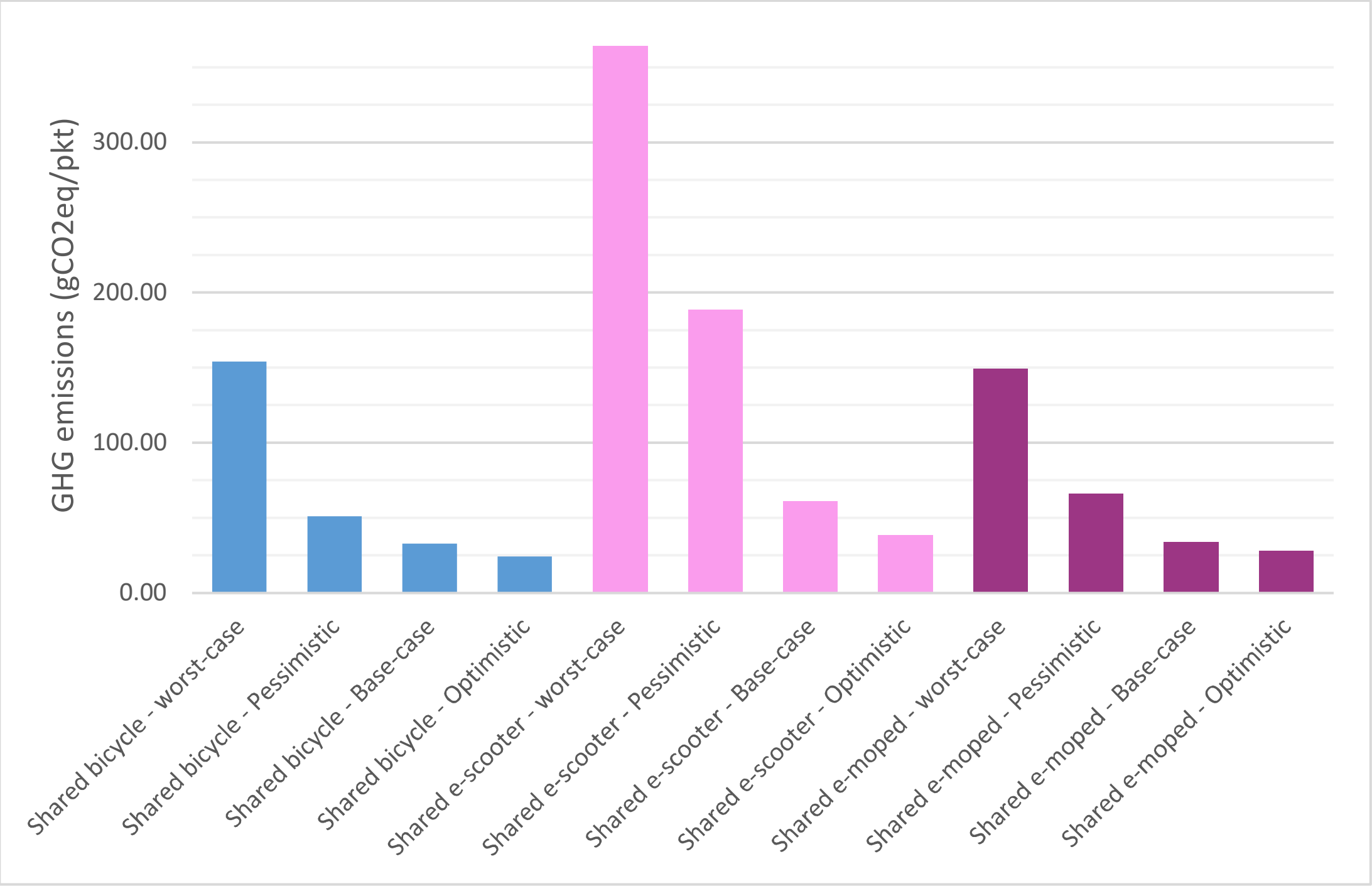

**Figure 4 Carbon footprint of the shared micromobility modes depending on the lifespan scenario**

The typical lifetime mileage considered for private bikes in the literature is 15 000 km [28]. In our study, it is updated based on French recent data. An additional simulation shows that considering a 15 000-km lifespan would increase the personal bike impacts by 20 to 25% on our five indicators. For instance, the carbon footprint rises from 12 to 15g of $CO_2eq$/km.

**Servicing distance**

Figure 5 presents the carbon footprint of the three shared modes depending on the servicing distance scenarios defined in section 2. Similar figures for the four other indicators are presented in the supplementary material. These results highlight the limited influence of servicing distances on the environmental ranking when they are traveled with e-vans in countries with a low-carbon intensity electricity. The environmental performance is particularly stable for bikes and mopeds, respectively limited to 10 and 15% between the worst-case and the optimistic scenarios, depending on the indicator. ES impact is more sensitive to the servicing distance, because the ES servicing distance also varies on a wider range. This is due to (a) the higher demand in servicing - for maintenance, rebalancing, but also charging, contrary to shared bikes only moved for maintenance and rebalancing - and (b) the variability of the battery type (swappable or not), contrary to the e-mopeds which are heavier and thus require a swappable battery. Between the worst-case (non-swappable battery and charging location far from the city center) and the optimistic scenarios (swappable battery and charging location in the city center), the carbon footprint varies by almost 30%, from 80g of $CO_2$eq/km to 58g of $CO_2$eq/km.

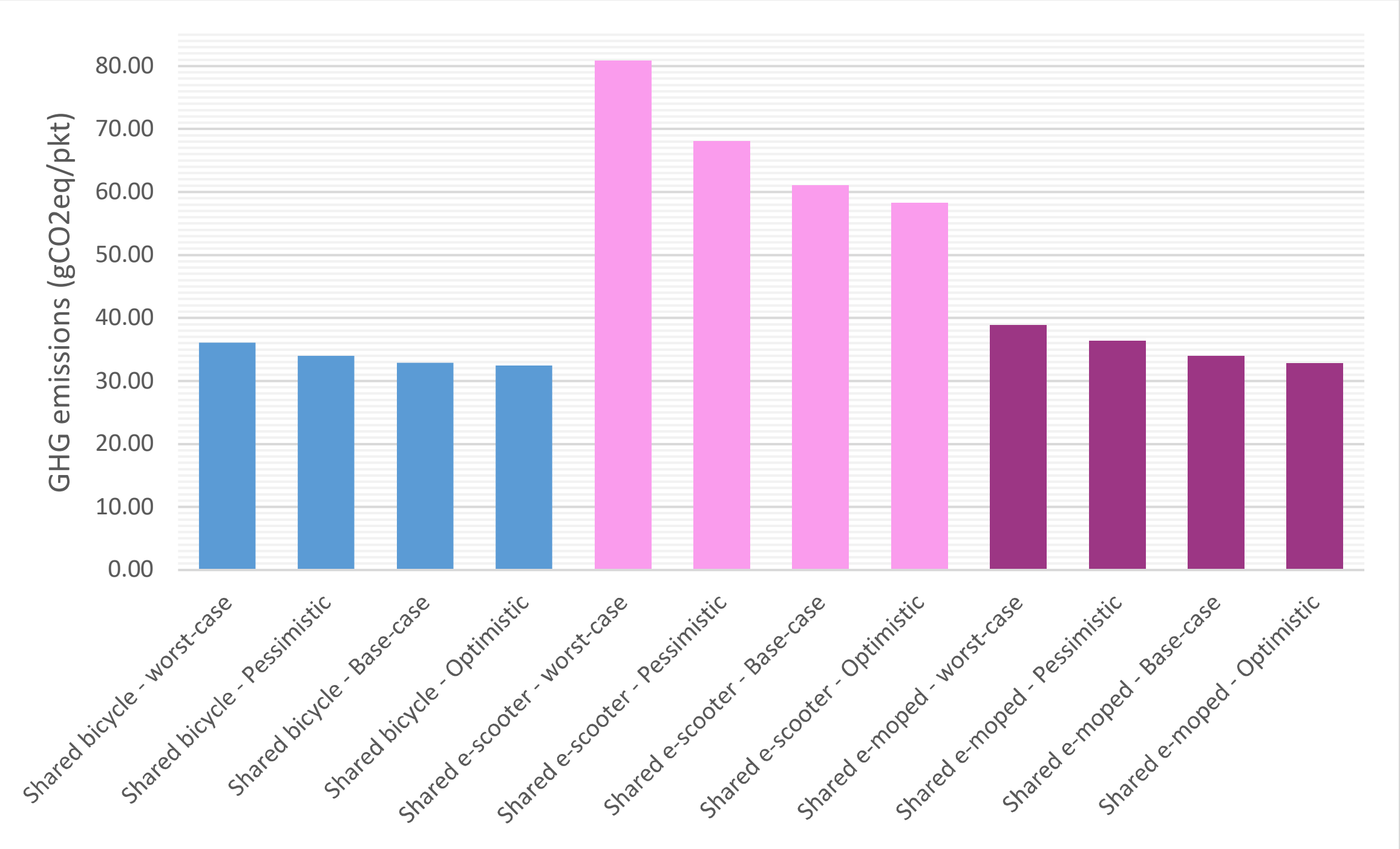

**Figure 5 Carbon footprint of the shared modes depending on the servicing distance**

**Shipping alternative**

No transportation is included in the vehicle LCIs from the assembly plant to the using location. Figure 6 presents the carbon footprint of the three shared modes depending on the servicing distance scenarios defined in section 2. Similar figures for the four other indicators are presented in the supplementary material. Results highlight that the shipping option is not very influential on the environmental performance of the shared modes if it is done by sea + road or rail + road (scenarios UE1, UE2, US1, US2, CN1, and CN2), whatever the environmental impact considered. For example, it increases the modal carbon footprints by 2 to 3% to send the vehicles from China to Europe by ship and truck (between +0.7 and 1.2g of $CO_2$eq/pkt). But air shipping strongly deteriorates the performance of the mode: for instance, it increases the carbon footprint of the shared bike mode by 57% in Europe (scenario EU3) and by 69% in the US (scenario US 3) compared to a sea + road shipping (resp. scenario UE1 and US1). In the case of the shared e-moped, the effect is even higher, with carbon footprints respectively increased by 79 and 96%.

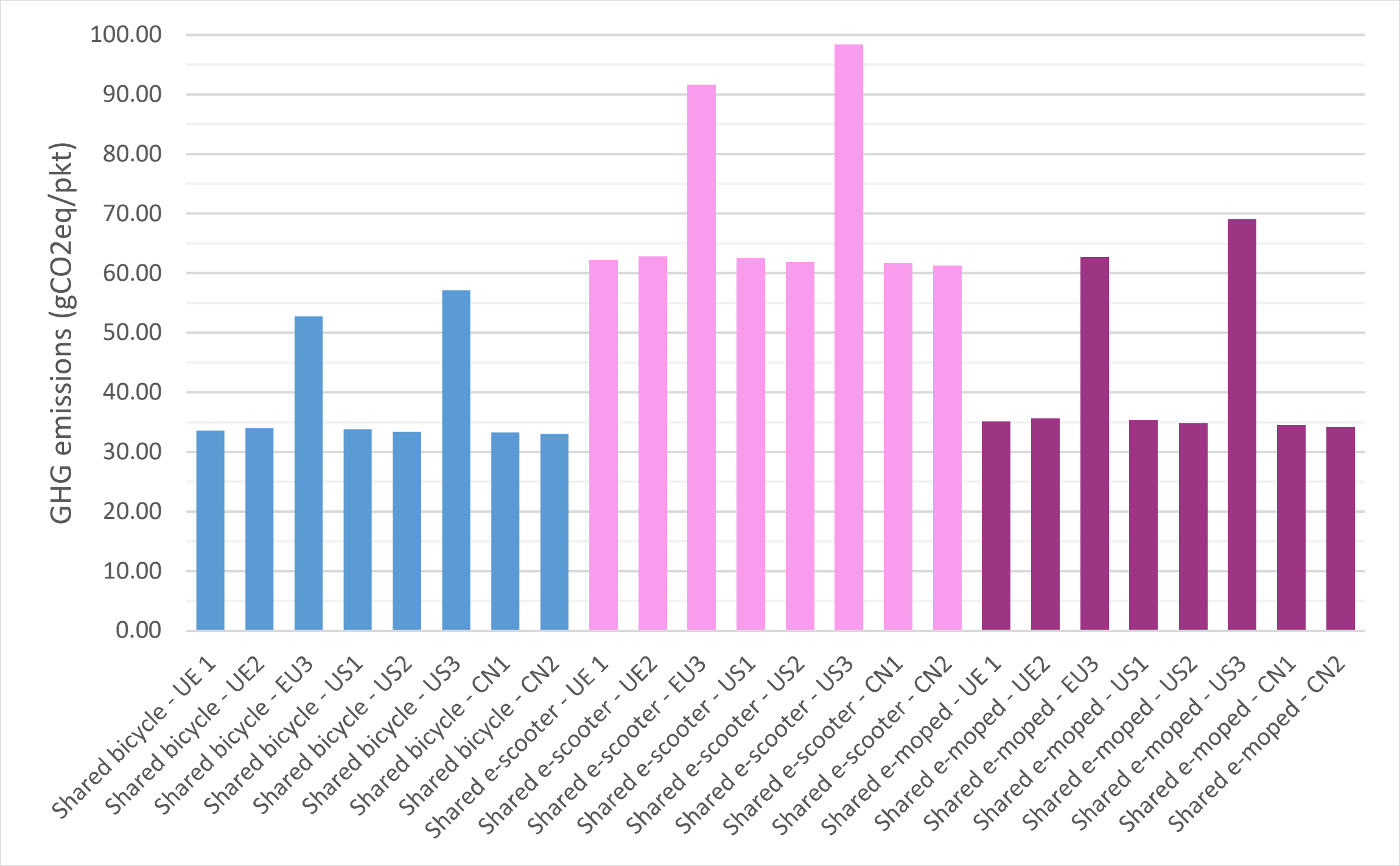

**Figure 6 Carbon footprint of the shared modes depending on shipping type**

**Electricity mix**

The electricity mix influences the impact of the use stage as well as the servicing stage, as electric vehicles are assumed to be used to maintain, balance, and charge the vehicles. Figure 7 presents the carbon footprint of the three shared modes depending on the electricity mix of twelve different countries, selected in section 2. Similar figures for the four other indicators are presented in the supplementary material. Results show that the environmental performances of shared bikes are not sensitive to the type of electricity mix on the five indicators. This is due to (1) the low electricity consumption of e-bikes and (2) the fact that the shared bike fleet is considered half electric and half mechanical, thus cutting the electricity consumption by a factor of two. Such a bike-sharing system emits 36g of $CO_2$eq/pkt in China - the country with the highest carbon intensity mix – while it emits 33g of $CO_2$eq/pkt in Norway – the country with the lowest carbon intensity mix. Under the baseline assumptions, the carbon footprint of docked shared bikes is 34.5g of CO2eq/km +/- 5% worldwide.

On the contrary, the environmental performance of shared ES and shared e-mopeds is highly influenced by the electricity mix. A shared ES emits around 60g of $CO_2$eq/pkt in Norway, Denmark, and France, when it can reach up to 92g of $CO_2$eq/pkt in China, i.e. 50% more emissions. Shared e-mopeds are even more sensitive: while they emit 32g of $CO_2$eq in Norway and Denmark, they reach 78g of $CO_2$eq/pkt in China, i.e. +144%. In countries with a high-carbon intensity electricity, the contribution of the use and servicing stages of ES can reach between 29% and 41% of the modal impact, respectively on DR and GWP in China (see supplementary material). In this country, the ES use stage contributes to 22% of the carbon footprint of the mode, and the servicing to 19% (with e-vans). On DHH, these contributions are the lowest, with the use stage representing 8% of the impact and the servicing 23%. For the shared e-moped, the use stage contributes to 52% of the carbon footprint of the mode, and the servicing to 11%. These contributions are also the lowest on DHH: 12% for the use stage and 8% for the servicing.

Under our assumptions, the three shared modes are in general ranked as follows, from the most to the less climate-friendly mode: bike > e-mopeds > ES. For instance, in the US, a shared bike emits 35g of $CO_2$eq/km, a shared e-moped 58g, and a shared ES 78g. But in countries with a low-carbon intensity electricity – like Norway, Denmark, and France, shared bikes and e-mopeds rank equally. This performance ranking is stable on DR and DE, as well as on EC in many countries – France, Spain, the Netherlands, the USA, the UK, Germany, Australia, and China. Nevertheless, ES are systematically slightly better than e-mopeds in terms of DHH, and bikes rank between ES and e-mopeds in Norway, Denmark, Canada, and Italy on EC.

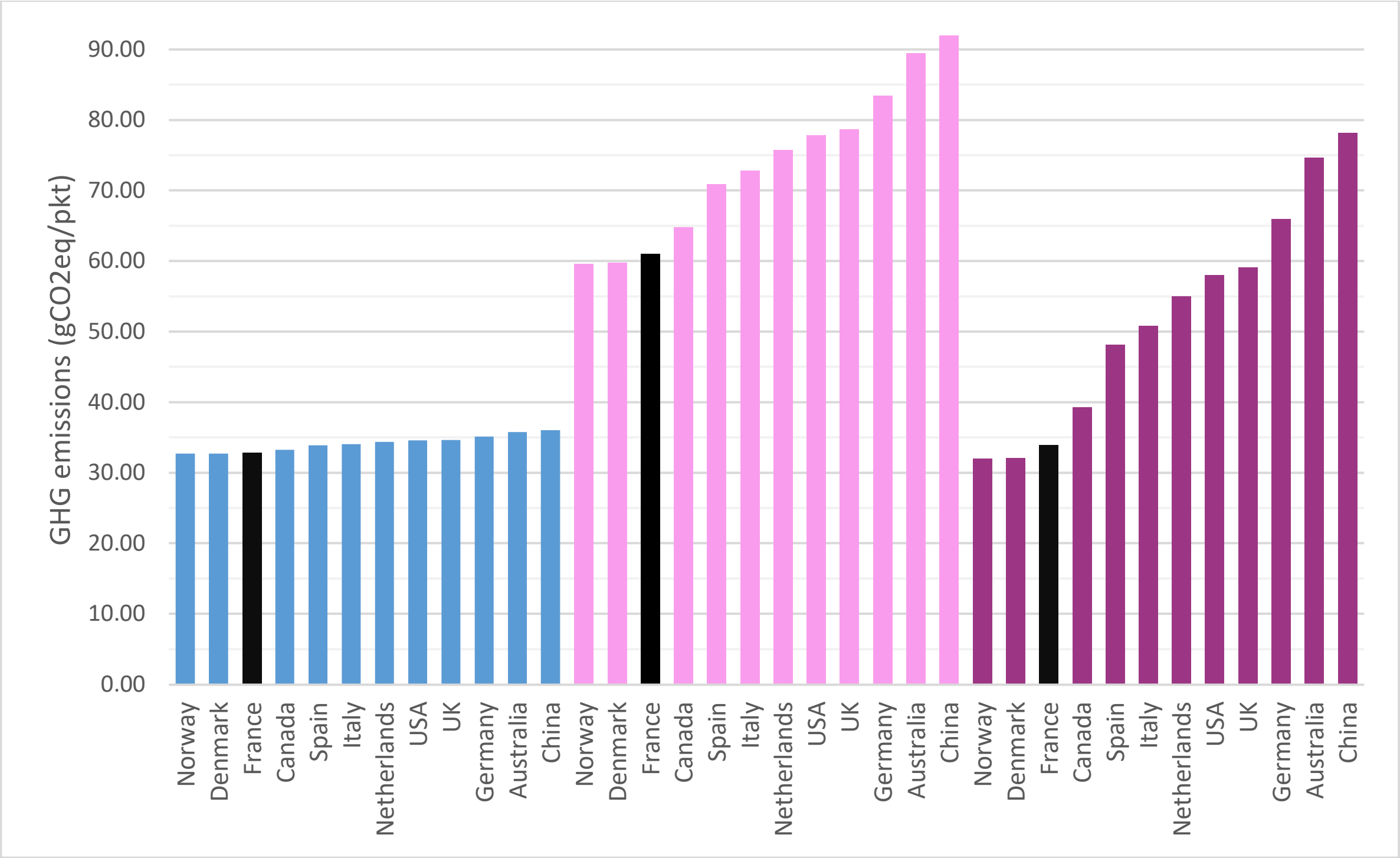

**Figure 7 Carbon footprint of the shared modes depending on the electricity mix: shared bikes in blue, shared ES in pink, shared e-mopeds in purple, and French reference in black**

# 6. Discussion and recommendations

## 6.1. Changing the paradigm in environmental assessments of transportation policies

The environmental contribution pattern of the different shared micromobility modes assessed in this article can be divided into two groups: electric modes and ICE modes. The performance of the ICE mode is driven by the use stage. On the contrary, the impact of electric micromobility is mainly driven by the vehicle component and generally by vehicle manufacturing. The use stage is only influential for shared ES and shared e-mopeds consuming high-carbon intensity electricity and on GWP, DR, and DE. It is only a major contribution (52%) in the country with the highest carbon intensity electricity (China) for shared e-mopeds. The servicing stage using e-vans is limited to 23% maximum of the total impact (human health damage in China). This must be considered when designing simplified environmental methodologies to support transportation decision-making: the classical approach, constrained to the use stage assessment, is not an acceptable option anymore and must be replaced by a full life cycle approach. For instance, eco-designing microvehicles to minimize their modal impact mainly requires to (1) optimize their manufacturing impact per pkt by choosing durable materials with low environmental impacts – e.g. steel instead of aluminum, (2) optimize the maintenance, and (3) maximize the recycling of each component/material at its highest-quality potential.

### 6.2. Infrastructure and servicing contributions

Infrastructure and servicing have their role to play in the environmental sustainability of micromobility modes. A previous study showed the marginal GWP contribution from the infrastructure for road modes in Paris [19]. Consistently, for private motorcycles, our study shows a 4% contribution from the pavement life cycle to the total carbon footprint. But this is especially true for GWP, and less for EC, DR, and DE where the infrastructure accounts for 12% of the motorcycle carbon footprint. Additionally, as the contribution of the vehicle diminishes with well-managed ES and bikes allowing for longer lifespans, the contribution from the infrastructure rises on the integrated modal impact. Moreover, specific infrastructure for shared modes like bike-sharing stations can have a substantial impact: the infrastructure contributes to 68% of the modal impact on EC for the Velib' bike. Optimizing bike-sharing stations is thus necessary to ensure the best environmental performance of the system, as operators already started to do. Nevertheless, the best time to replace existing high-consuming stations must be assessed, as there is a trade-off to find between the environmental impact of a new station production and electricity savings during the use stage. The same considerations apply to wisely replace all kinds of preceding technologies with new technologies under an eco-friendly orientation [50].

The servicing stage, i.e. the fleet management, is also influential on the environmental performance of shared microvehicles. It was estimated to be responsible for half of the carbon footprint of Parisian and US first generation shared ES [19, 26]. Our results show lower contributions because operators mostly switched to e-vans. Some operators also opted for swappable batteries, allowing to reduce the size of the servicing vehicles, thus the kilometers traveled to charge one vehicle and the servicing impact. While the second option is indubitably environmentally virtuous as far as the fleet replacement has been well driven, the first option is questionable on a multicriteria approach, as previous comparisons between electric and ICE vehicles showed non-unanimous rankings amongst indicators in different national contexts [51].

### 6.3. Carbon performance of micromobility versus other modes

**Paris's case**

Figure 8 shows the carbon footprint of most of the Parisian transportation modes in 2020, including the modes assessed in this study and other Parisian modes assessed with the same methodology in another study [19]. Carbon footprints are ranked from the less impacting – walking, which emits 2g of $CO_2$eq/km - to the most emitting - taxis and ride-hailing, accounting for 300g of $CO_2$eq/pkt. It highlights how electric micromobility ranks between ICE modes and electric public transportation/active modes in Paris on GWP. Actives modes – i.e. walking and cycling - are the less emitting modes with the metro and the RER train (light rail). Nevertheless, the shared bike is ranked sixth due to its shorter lifetime mileage compared to the personal bike, but also due to the impact of bike-sharing stations and servicing.

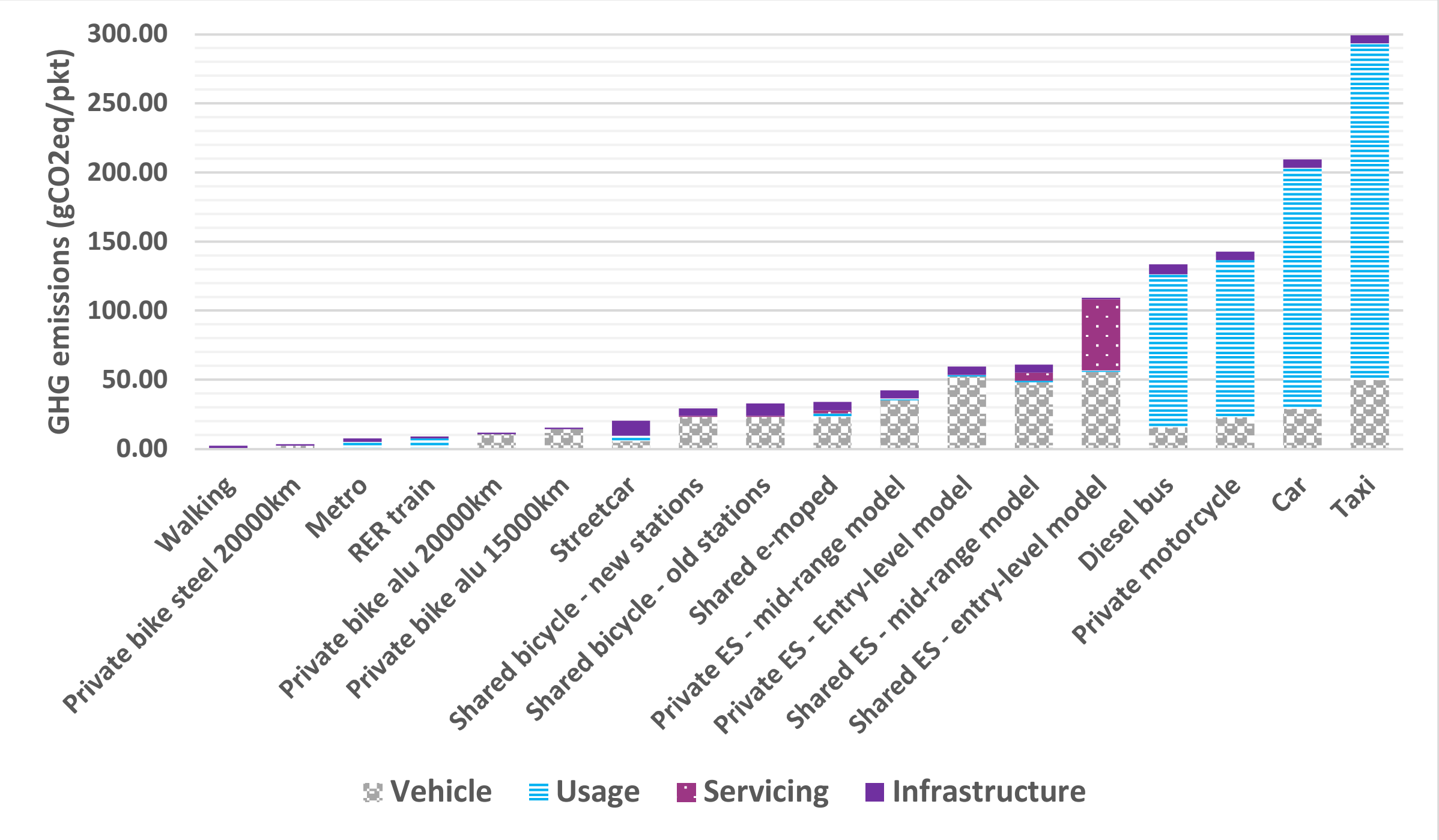

**Figure 8 Carbon footprint comparison of different transportation modes in Paris per pkt**

**US conditions**

To verify this carbon ranking in other conditions, we compare the carbon footprints of the shared modes calculated with the average US electricity mix (subsection 5.3) to the performances calculated by Chester and Horvath [32] in US conditions for public transportation and personal cars as well as the impact of walking and cycling with a personal bike calculated in this study (in French conditions but quite similar to the US for these modes). Results are presented in Figure 9 and show that, in the US, shared micromobility modes globally rank between active modes and public transportation/private car modes on GWP. Public transportation modes present higher carbon footprints in US conditions than in Paris mainly due to a higher carbon intensity electricity and a lower vehicle occupancy rate. But the carbon footprint from the infrastructure is also higher in the US. This can be explained by a more intensive infrastructure usage in Paris, but also potentially by underestimated LCIs in the ecoinvent processes used in the Parisian study.

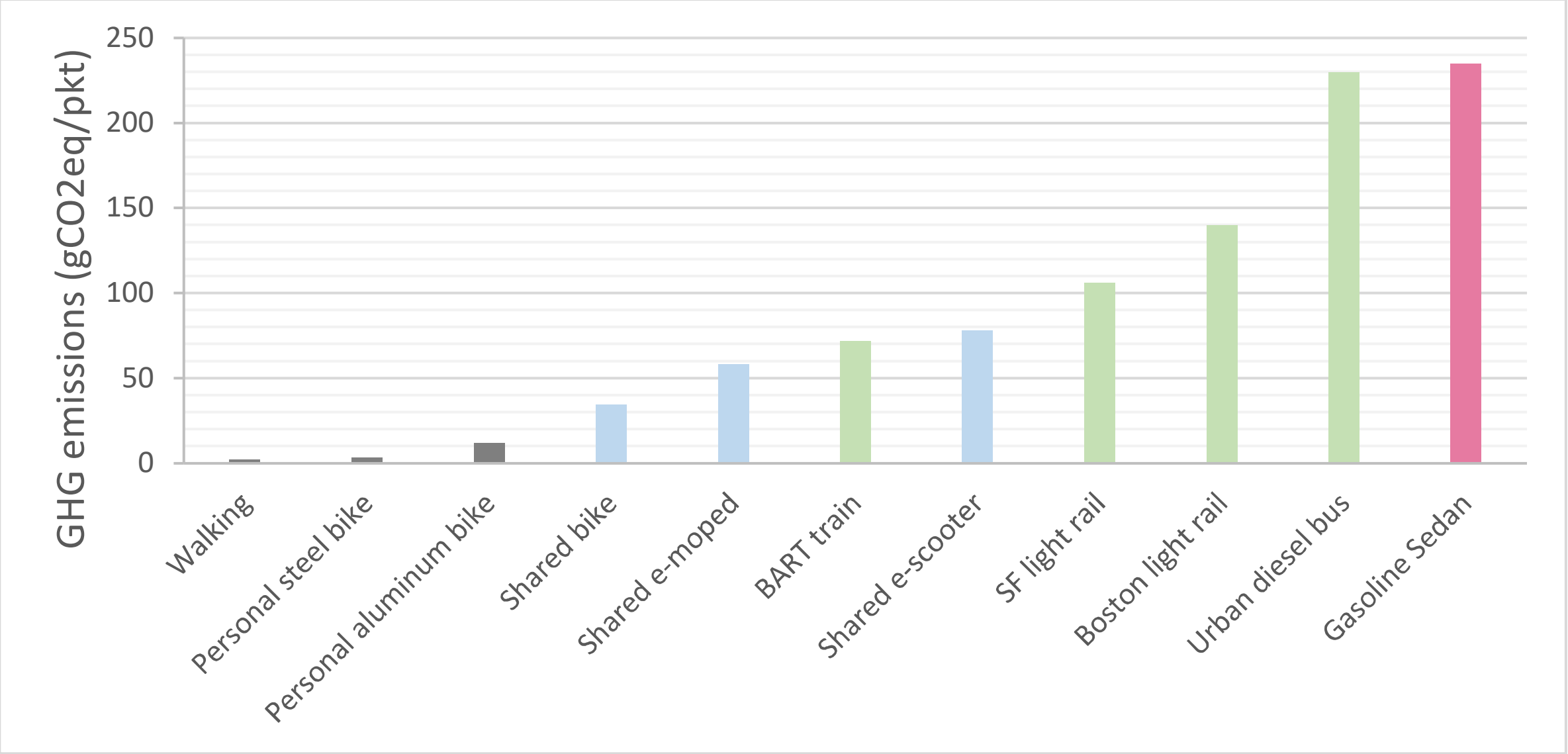

**Figure 9 Carbon footprint of US modes per passenger-kilometer traveled - in grey active modes, in blue shared micromobility modes, in green public transportation and in red, private car**

## 7. Conclusion

The environmental performance of micromobility is investigated within a large set of conditions, under country-specific and non-country specific scenario analyses. Results show that personal micromobility globally ranks better than shared micromobility due to higher vehicle lifespans, but lifetime mileages need to be investigated further. Generally, the shared bike ranks better than the shared e-moped, itself ranking better than the shared ES. Servicing is influential on shared ES performance. Vehicle shipping does not affect the performance of the mode, unless done by aircraft. The electricity mix impacts ES and e-moped performances except in countries with very low-carbon intensity electricity. Electric micromobility globally ranks between active modes and personal ICE modes.

Moreover, our analysis highlights the importance of multicriteria integrated modal LCA to properly capture environmental key performance factors and design eco-friendly transportation policies considering potential burden-shifting and how they could jeopardize human well-being and planet preservation. Considering such a system boundary and a complete scope of indicators – for instance, the three endpoint indicators - is a change of paradigm compared to the most popular environmental assessments. Indeed, they are, first, restrained to GHG emissions considerations, and, second, often limited to the use stage. Such an approach was an acceptable simplification to compare personal ICE modes but has become obsolete with the recent changes in mobility technologies, services, and behaviors.

**Declarations**

None

**Availability of data and material**

The authors declare that they have no competing interests.

**Funding**

No funding has been received for this study.

**Acknowledgments**

We want to thank the few brave operators and suppliers of shared mobility systems which accepted to share their private data to conduct an environmental assessment of quality: Cityscoot, Smove, DOTT, and JC Decaux.

# SUPPLEMENTARY MATERIAL

# Environmental performance of shared modes and personal alternatives using integrated LCA

A. de Bortoli

## Table of contents

## 7.1. List of tables

## 7.2. List of figures

Nota bene: 2.09E+02 is the scientific notation of a number and must be read as $2.09 \times 10^2 = 209$

# 8. Infrastructure impact calculation

### Infrastructural number of units and LCIs

Paris' infrastructure surfaces are reported by Breteau (2016). The pavement network in Paris is composed of 1220 hectares of pavement and 198 hectares of street parking lots. The cycle lanes surface measures 106 hectares, and 7 more hectares are devoted to bike stations and parking lots.

The LCIs of the different types of infrastructures are specified in Table 6 and come from de Bortoli and Christoforou (2020)Table 9. The gravel volumetric mass is supposed equal to 1.54 t/m$^3$ and the hot mix asphalt volumetric mass to 2.3 t/m$^3$ after compaction.

Table 6 LCIs of the different types of infrastructures in Paris

| | Cycle lane | Pavement | Sidewalk |
|---|---|---|---|
| Functional Unit | 1 linear meter/year | 1 m²/ year | 1 m²/y |
| Binder (kg) | 0.246675 | 0.705333 | 0.2875 |
| Gravel (kg) | 9.370075 | 17.71133 | 6.4375 |
| Concrete block (kg) | 5.4 | 0 | 0 |
| Truck transportation (tkm) | 0.7508375 | 0.920833 | 0.33625 |
| HMA manufacturing (kg) | 4.11125 | 11.07056 | 2.875 |

The shared e-scooters and personal bikes use sidewalks and previous car parking lots to park in Paris: we will neglect this impact because of a lack of data. However, it must be extremely low, like the contribution of the infrastructure (de Bortoli and Christoforou 2020). Nevertheless, the impact of the shared bike station will be estimated to check the contribution of its components (see LCIs in Table 7 and Table 8) and the electricity consumption.

**Table 7 LCIs for one shared bike dock**

| Process | Quantity | Unit |
|---|---|---|
| steel, low-alloyed, hot rolled \| market for steel, low-alloyed, hot rolled - GLO | 23.0 | kg |
| polypropylene, granulate \| market for polypropylene, granulate - GLO | 0.5 | kg |
| stretch blow moulding \| market for stretch blow moulding - GLO | 0.5 | kg |

**Table 8 LCIs to produce one shared bike e-kiosk**

| **Materials (for 2019)** | Ecoinvent process | One station | unit |
|---|---|---|---|
| Aluminum | market for aluminum, primary ingot, GLO | 125.6 | kg |
| | Market for sheet rolling, aluminium, GLO | 125.6 | kg |
| Steel (inside components + stainless steel + wires) | Market for steel, low-alloyed, GLO | 116.6 | kg |
| | Market for sheet rolling, steel, GLO | 116.6 | kg |
| Foam | Market for polyurethane, flexible foam, GLO | 0.3 | kg |
| Plastic | Market for polypropylene, granulate, GLO | 17 | kg |
| | Market for blow moulding, GLO | 17 | kg |
| Rubber | Market for synthetic rubber, GLO | 8.5 | kg |
| Screen and printed board | Market for computer, laptop | 1 | unit |

# 9. Vehicle life Cycle Inventory details

**Table 9 LCIs for different vehicles life stage, except the use stage**

<table>
<tr><th>Stage</th><th>Production</th><th></th><th>Maintenance</th><th></th><th>EoL</th><th></th></tr>
<tr><td>Mobility macro-process</td><td>Process</td><td>Quantity</td><td>Process</td><td>Quantity</td><td>Process</td><td>Quantity</td></tr>
<tr><td>Personal bicycle</td><td>"market for bicycle, GLO"</td><td>1 item(s)</td><td>"market for maintenance, bicycle, GLO"</td><td>20.6/17 item(s)</td><td colspan="2">Included in the production</td></tr>
<tr><td rowspan="3">Shared e-bicycle</td><td>"market for bicycle GLO"</td><td>20.6/17 item(s)</td><td>"market for maintenance, bicycle, GLO"</td><td>20.6/17 item(s)</td><td colspan="2">Included in the production</td></tr>
<tr><td>"electric motor, vehicle"</td><td>4 kg</td><td colspan="2">No maintenance</td><td>No process</td><td></td></tr>
<tr><td>“battery, Li-ion, rechargeable, prismatic, GLO”</td><td>2.4 kg</td><td colspan="2">No maintenance</td><td>"market for used Li-ion battery, GLO"</td><td>2.4 kg</td></tr>
<tr><td rowspan="2">Shared e-moped</td><td>"market for electric scooter, without battery, GLO"</td><td>98 kg</td><td>"market for maintenance, electric scooter, without battery, GLO</td><td>1 item(s)</td><td colspan="2">Included in the production</td></tr>
<tr><td>"market for battery cell, Li-ion, GLO"</td><td>28 kg</td><td colspan="2">No maintenance</td><td>"market for used Li-ion battery, GLO"</td><td>-28 kg</td></tr>
<tr><td>Personal motorcycle</td><td>market for motor scooter, 50 cubic cm engine, GLO</td><td>127kg/90kg= 1.4 item(s)</td><td>"market for maintenance, motor scooter, GLO</td><td>1.4 item(s)</td><td colspan="2">Included in the production</td></tr>
</table>

**Table 10 LCI for one new generation FFES, except for the use stage**

| Flow | Amount | Unit |
|---|---|---|
| aluminium alloy, AlMg3 \| market for aluminium alloy, AlMg3, GLO | 10.628 | kg |
| aluminium, cast alloy \| market for aluminium, cast alloy, GLO | 0.475 | kg |
| battery cell, Li-ion \| battery cell production, Li-ion, CN | 2.318 | kg |
| charger, for electric scooter \| charger production, for electric scooter, GLO | 0,385 | kg |
| electric motor, for electric scooter \| electric motor production, for electric scooter, GLO | 2.201 | kg |
| electricity, medium voltage, aluminium industry \| market for electricity, medium voltage, aluminium industry, CN | 12.777 | kWh |
| heat, central or small-scale, other than natural gas \| market for heat, central or small-scale, other than natural gas, ROW | 0.358 | MJ |
| heat, district or industrial, natural gas \| market for heat, district or industrial, natural gas, ROW | 25.22 | MJ |
| light emitting diode \| light emitting diode production, GLO | 0,016 | kg |
| municipal solid waste \| market for municipal solid waste, ROW | -8.345 | kg |
| polycarbonate \| market for polycarbonate, GLO | 0.508 | kg |
| powder coat, aluminium sheet \| market for powder coat, aluminium sheet, GLO | 0.649 | m2 |
| printed wiring board, surface mounted, unspecified, Pb containing \| printed wiring board production, surface mounted, unspecified, Pb containing, GLO | 0.059 | kg |
| steel, low-alloyed \| market for steel, low-alloyed, GLO | 2.502 | kg |
| synthetic rubber \| market for synthetic rubber, GLO | 2.198 | kg |
| tap water \| market for tap water, ROW | 1.38 | kg |
| transistor, wired, small size, through-hole mounting \| market for transistor, wired, small size, through-hole mounting, GLO | 0.062 | kg |
| transport, freight train \| market for transport, freight train, CN | 106.958 | t*km |
| transport, freight train \| transport, freight train, diesel, CN | 57.044 | t*km |
| transport, freight train \| transport, freight train, electricity, ROW | 35.653 | t*km |
| transport, freight, lorry, unspecified \| transport, freight, lorry, all sizes, EURO5 to generic market for transport, freight, lorry, unspecified, ROW | 35.653 | t*km |
| used electric bicycle \| market for used electric bicycle, GLO | -1.08 | Item(s) |
| used Li-ion battery \| market for used Li-ion battery, GLO | -2.318 | kg |
| wastewater, average \| market for wastewater, average, GLO | -0.001 | m3 |
| welding, arc, aluminium \| market for welding, arc, aluminium, GLO | 1.391 | m |

**Table 11 Tail pipe regulated emissions (CO, CO2, HC, NOx and PM) and fuel consumption of an average motorcycle in France in 2020 according to the HBEFA V4.1 calculator**

| Pollutant | Selected ecoinvent flow | Emission category | Emission factor [g/Vehkm] |
|---|---|---|---|
| CO | Carbon monoxide, fossil (emission to air/high population density) | hot | 2.09E+00 |
| CO2 | Carbon dioxide, fossil (emission to air/high population density) | hot | 8.13E+01 |
| FC | market for petrol, low-sulfur, Europe without Switzerland | hot | 2.56E+01 |
| HC | Hydrocarbons, aliphatic, alkanes, unspecified (emission to air/high population density) | evap | 2.81E-01 |
| HC | Hydrocarbons, aliphatic, alkanes, unspecified (emission to air/high population density) | hot | 2.75E-01 |
| NOx | Nitrogen oxides, FR (emission to air/high population density) | hot | 7.93E-02 |
| PM | Particulates, < 2.5 um (emission to air/high population density) | hot | 1.91E-02 |

Calculator accessible here : https://www.hbefa.net/Tools/EN/MainSite.asp

## 10. Lifetime mileage estimate of private bikes

For now on, when no data were available on the lifetime mileage of private bikes, the 15 000 km figure proposed by Luenberger and Frischknecht was typically used (2010). Nevertheless, this figure was not justified on field data. A recent survey on users of bikes and e-scooters in big French cities reviewed some characteristics of these two private vehicles in French households in 2020 (6t-bureau de recherche 2020). A question was dedicated to knowing when the respondent's bike was bought. From the responses, we extrapolated an average bike's age (Table 7). By multiplying the average lifetime of the bikes and the percent of respondents and by summing the sub-products, we get 5.4 years, equal to the average age of the French personal bikes. We can consider that twice that amount, i.e. 10.8 years, is a good estimate of the French personal bike lifespan. It is far above the 4 to 6 years estimated in China (Sun and Ertz 2021).

**Table 12 Answer to the question "when the bike was bought", percent of respondents, and estimates of average lifetime**

| Answer | Percent of respondents (%) | Average lifetime (months) | Average lifetime (years) |
|---|---|---|---|
| Less than 2 months ago | 3 | 1 | 0.083 |
| 2 to 6 month ago | 5 | 4 | 0.33 |
| 6 months to 1 year ago | 8 | 9 | 0.75 |
| 1 to 2 years ago | 17 | 18 | 1.5 |
| 3 to 5 years ago | 27 | 48 | 4.0 |
| 6 to 10 years ago | 20 | 96 | 8.0 |
| More than 10 years ago | 20 | 180 | 12 |

Secondly, another question was asked about the bike usage frequency, and the researchers who conducted the survey estimated the frequency of use and the bike use time per day (Table 13). From these figures, we can calculate the annual mileage for each frequency of use and multiply it by the percent of respondents. By summing the sub-products, we obtain 1854 km, the average annual distance traveled by a personal bike in France.

**Table 13 Answer to the question "how often do you use your bike" and indicators extrapolation**

| Answer | Percent of respondents (%) | Bike usage frequency (/week) | Use time (h/day) | Use time (h/year) | Annual mileage (km) |
|---|---|---|---|---|---|
| Everyday or almost | 17 | 7 | 1 | 365 | 5475 |
| 2-3 times a week | 26 | 3 | 0.43 | 157 | 2354 |
| Once a week | 23 | 1 | 0.14 | 51.1 | 766.5 |
| 2-3 times a month | 21 | 0.7 | 0.1 | 36.5 | 547.5 |
| Once a month | 12 | 0.2 | 0.03 | 11.0 | 164.2 |

By multiplying the average lifespan (10.8 years) with the average annual distance traveled (1854 km), we can estimate a personal bike in France has a lifetime mileage of 20 074 km, that we will round down at 20 000 km. To our knowledge, this figure is the first proper national estimate of a personal bike lifetime mileage.

# 11. Life cycle stage and component contributions by indicator

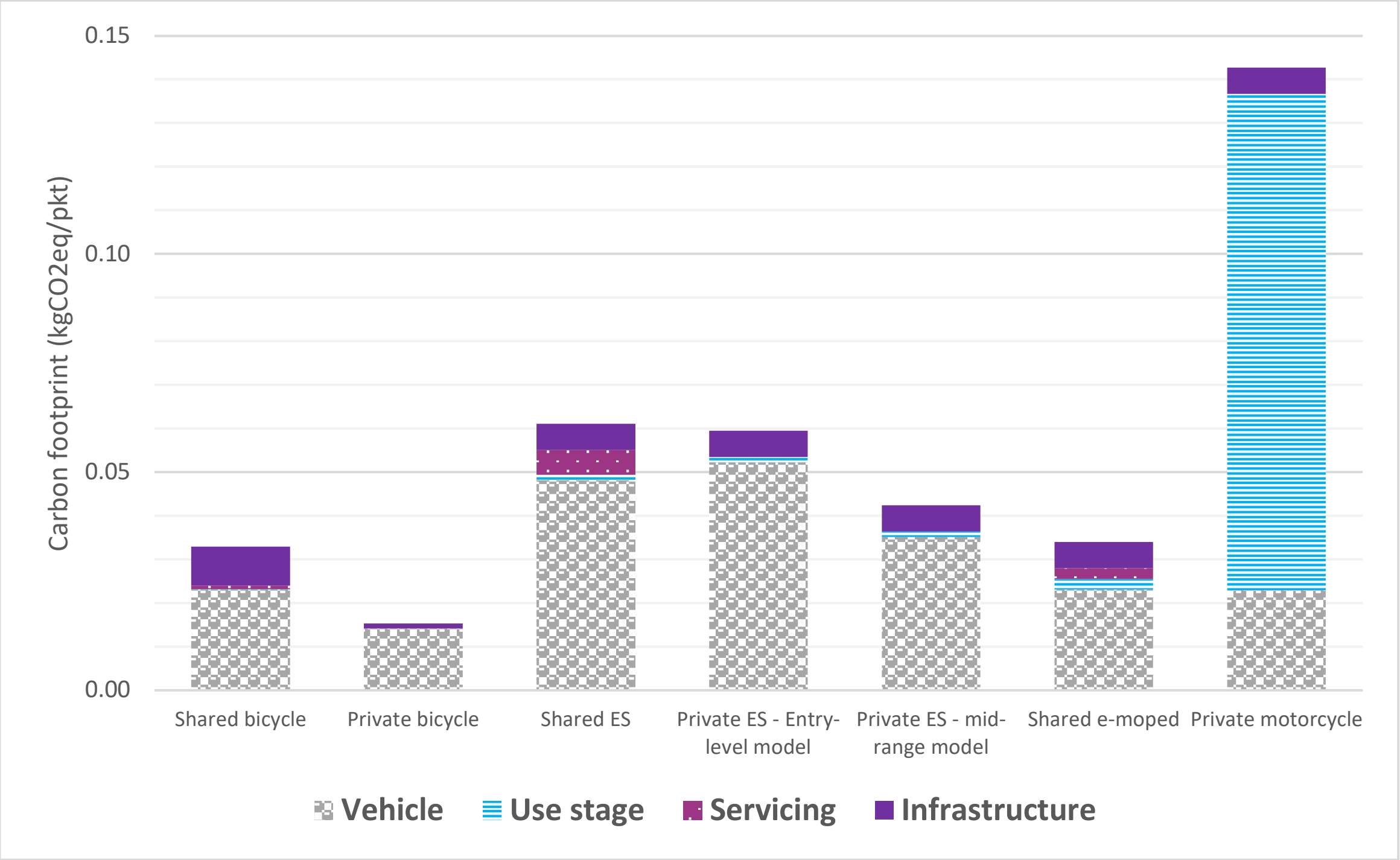

**Figure 10 Impacts and contributions to the Global Warming Potential by mode**

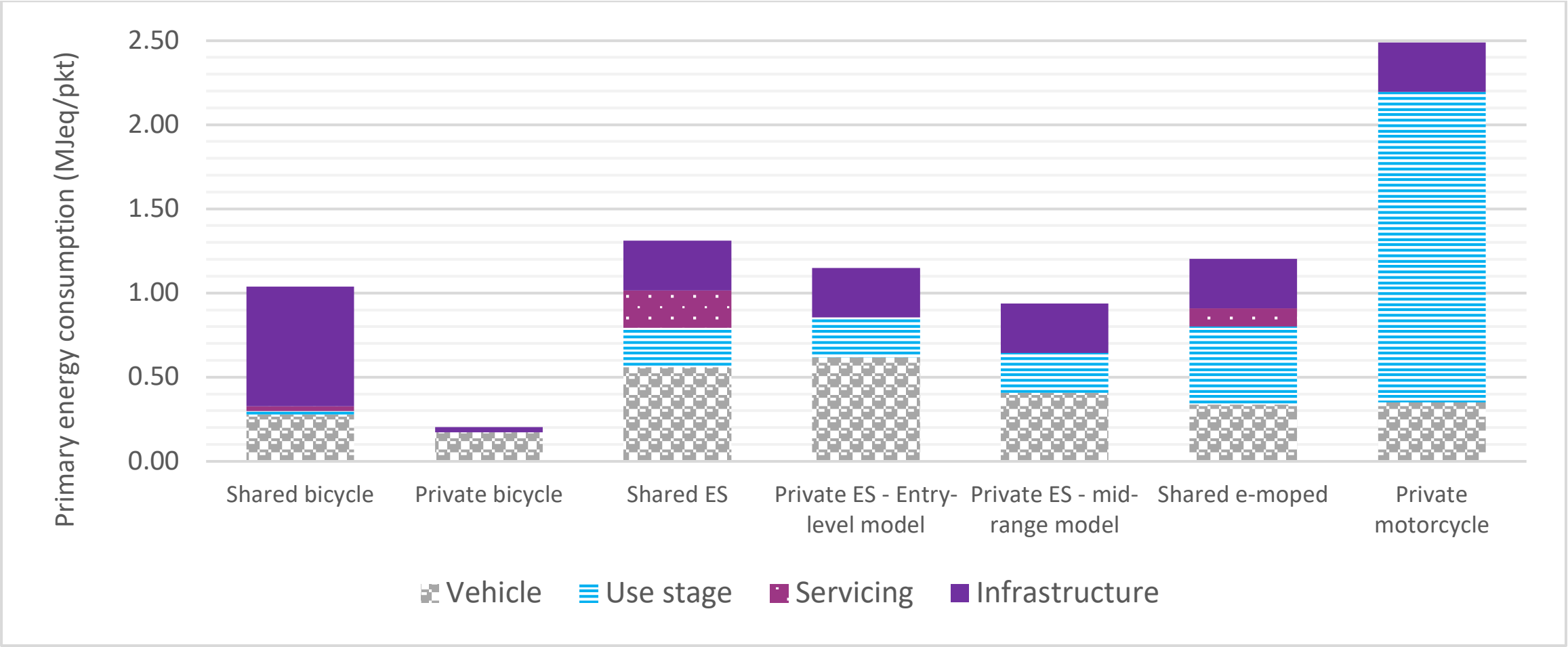

**Figure 11 Impacts and contributions to the primary energy consumption by mode**

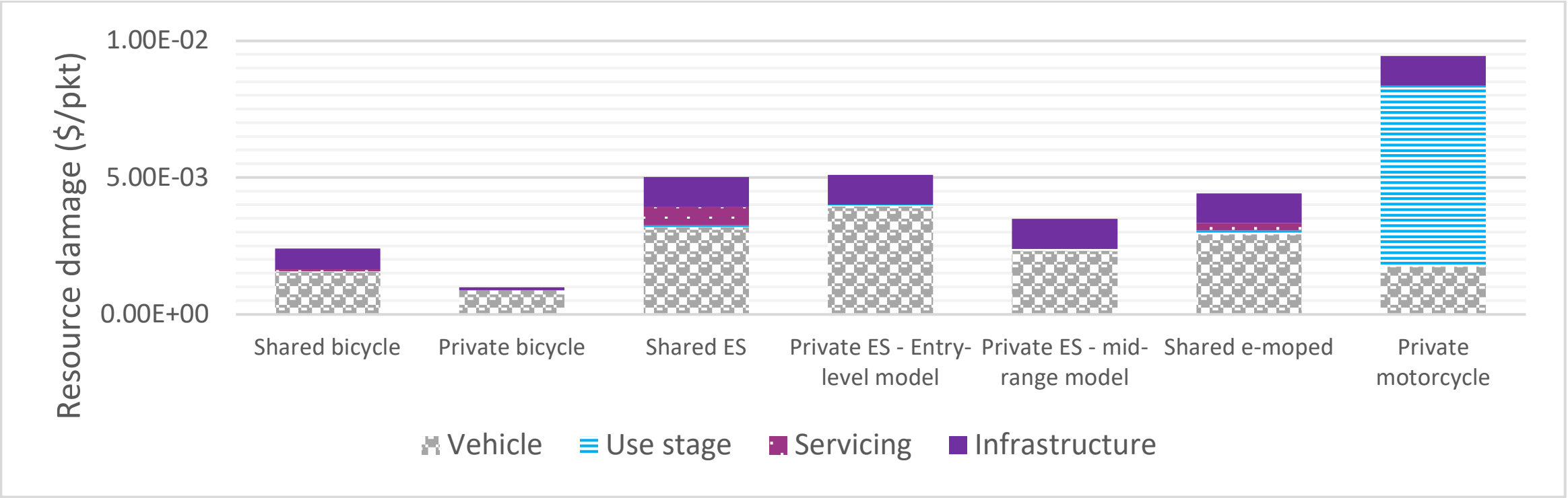

**Figure 12 Impacts and contributions to the resource damage by mode**

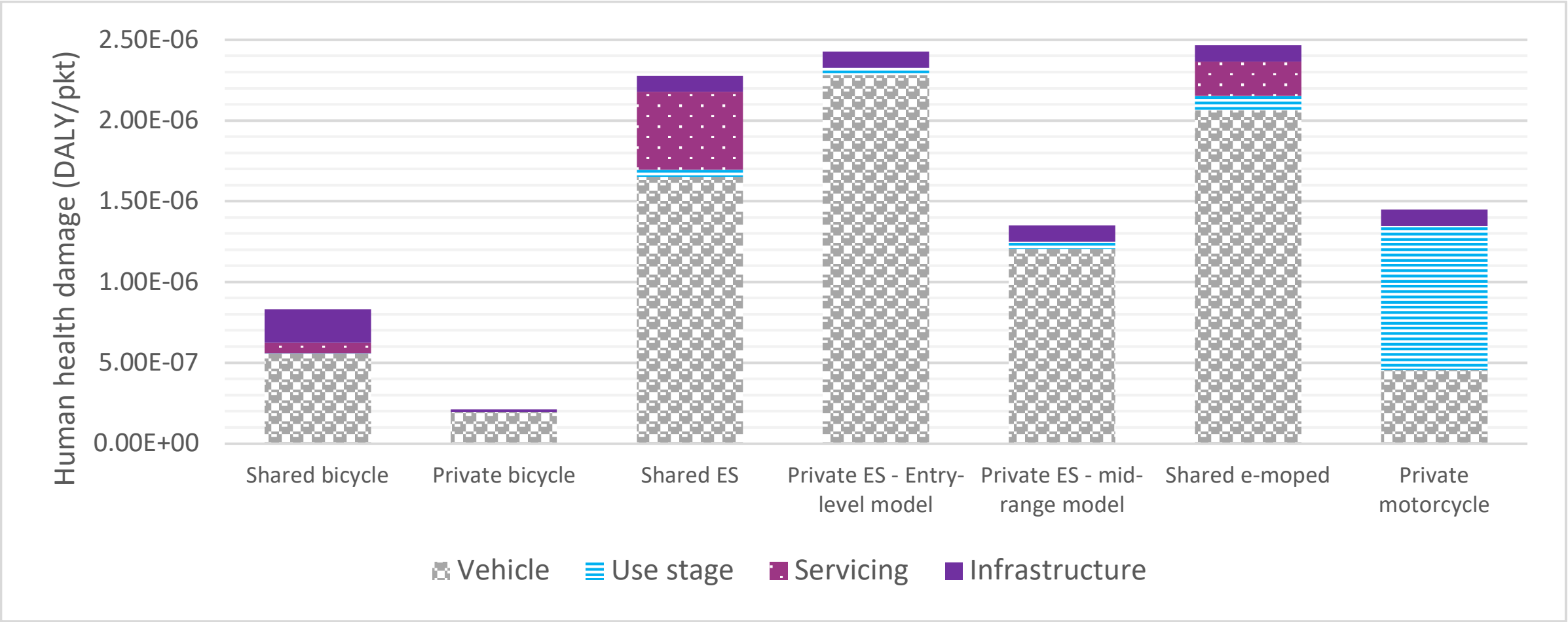

**Figure 13 Impacts and contributions to the human health damage by mode**

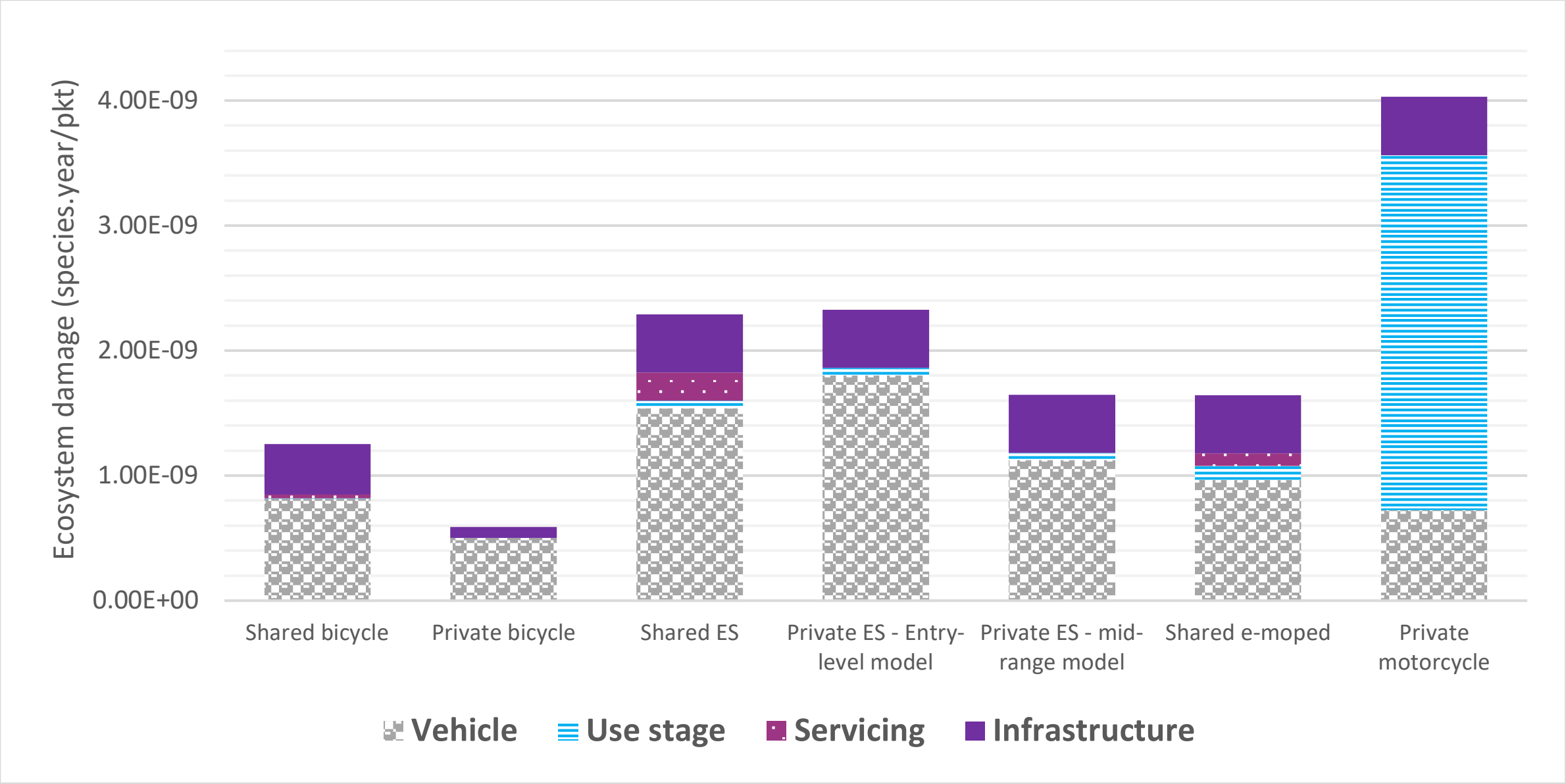

**Figure 14 Impacts and contributions to the ecosystem damage by mode**

# 12. Complementary results from scenario analyses

In these scenario analyses, the histograms show shared e-bikes environmental performances in blue, shared e-scooters performances in pink, and shared e-mopeds in purple. The case of France can be highlighted in black.

**Lifespan scenario analyses**

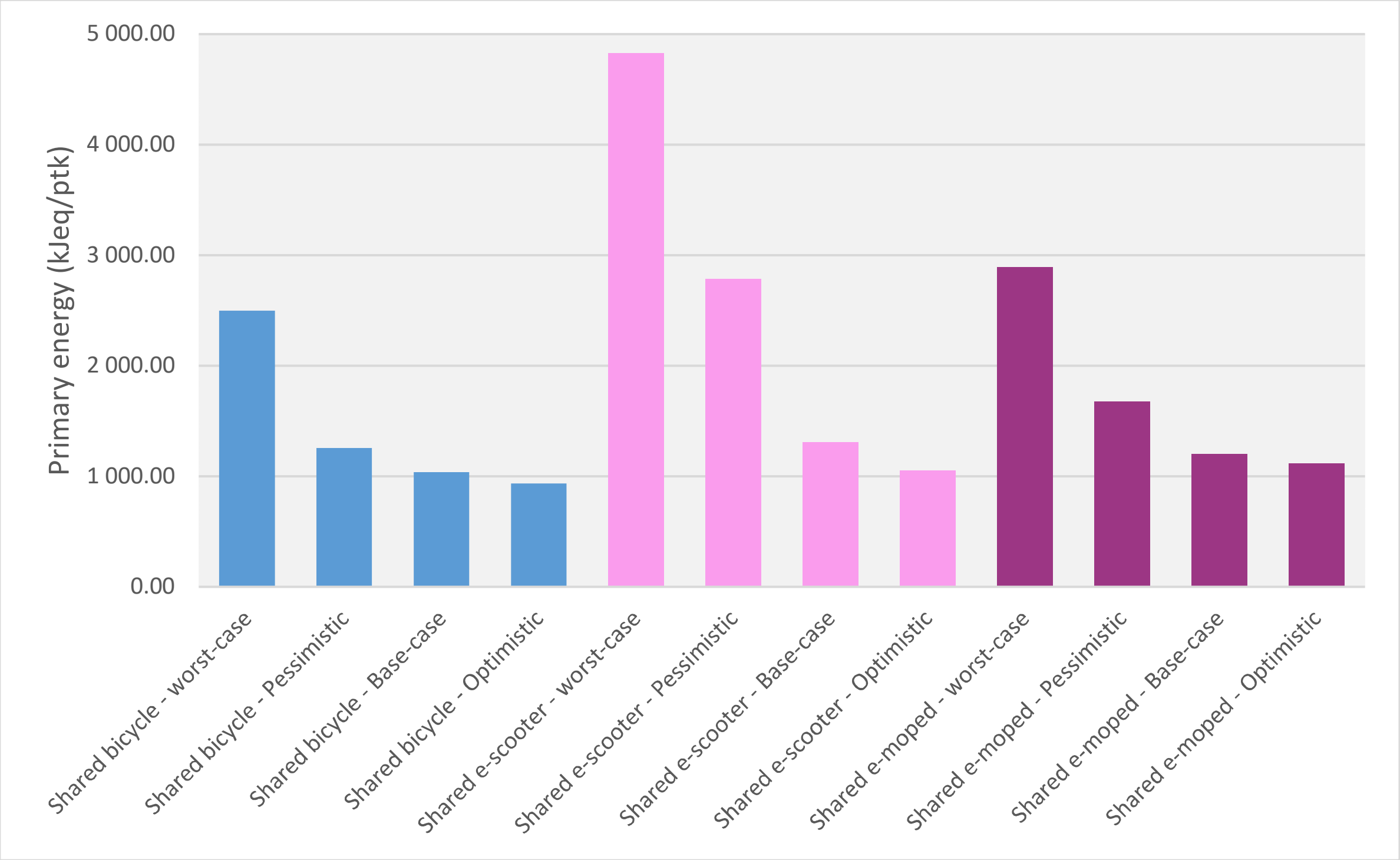

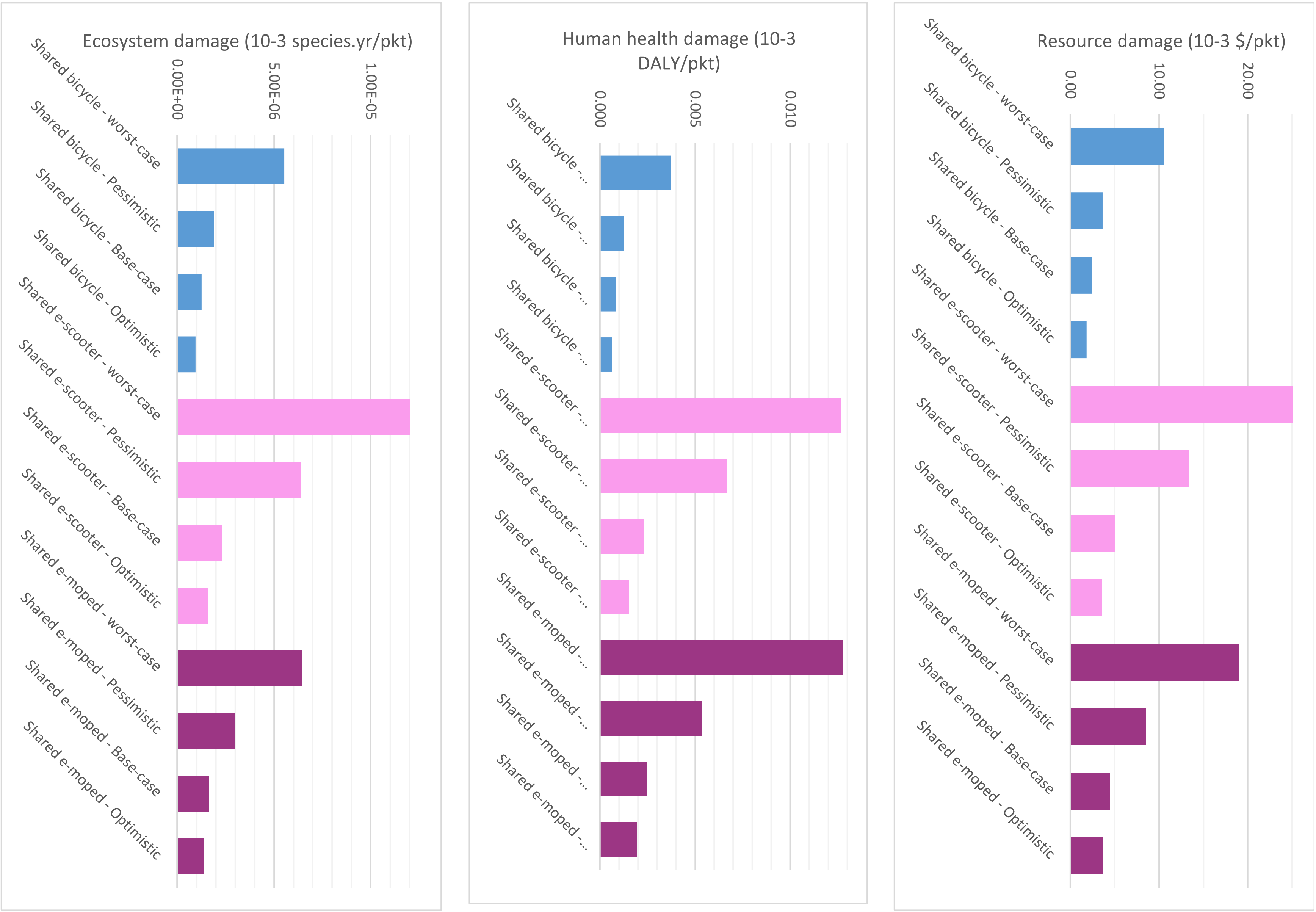
Ecosystem damage (10-3 species.yr/pkt)
0.00E+00
5.00E-06
1.00E-05
Shared bicycle - worst-case
Shared bicycle - Pessimistic
Shared bicycle - Base-case
Shared bicycle - Optimistic
Shared e-scooter - worst-case
Shared e-scooter - Pessimistic
Shared e-scooter - Base-case
Shared e-scooter - Optimistic
Shared e-moped - worst-case
Shared e-moped - Pessimistic
Shared e-moped - Base-case
Shared e-moped - Optimistic
Human health damage (10-3 DALY/pkt)
0.000
0.005
0.010
Shared bicycle -...
Shared bicycle -...
Shared bicycle -...
Shared bicycle -...
Shared e-scooter -...
Shared e-scooter -...
Shared e-scooter -...
Shared e-scooter -...
Shared e-moped -...
Shared e-moped -...
Shared e-moped -...
Shared e-moped -...
Resource damage (10-3 $/pkt)
0.00
10.00
20.00
Shared bicycle - worst-case
Shared bicycle - Pessimistic
Shared bicycle - Base-case
Shared bicycle - Optimistic
Shared e-scooter - worst-case
Shared e-scooter - Pessimistic
Shared e-scooter - Base-case
Shared e-scooter - Optimistic
Shared e-moped - worst-case
Shared e-moped - Pessimistic
Shared e-moped - Base-case
Shared e-moped - Optimistic

**Servicing distance scenarios**

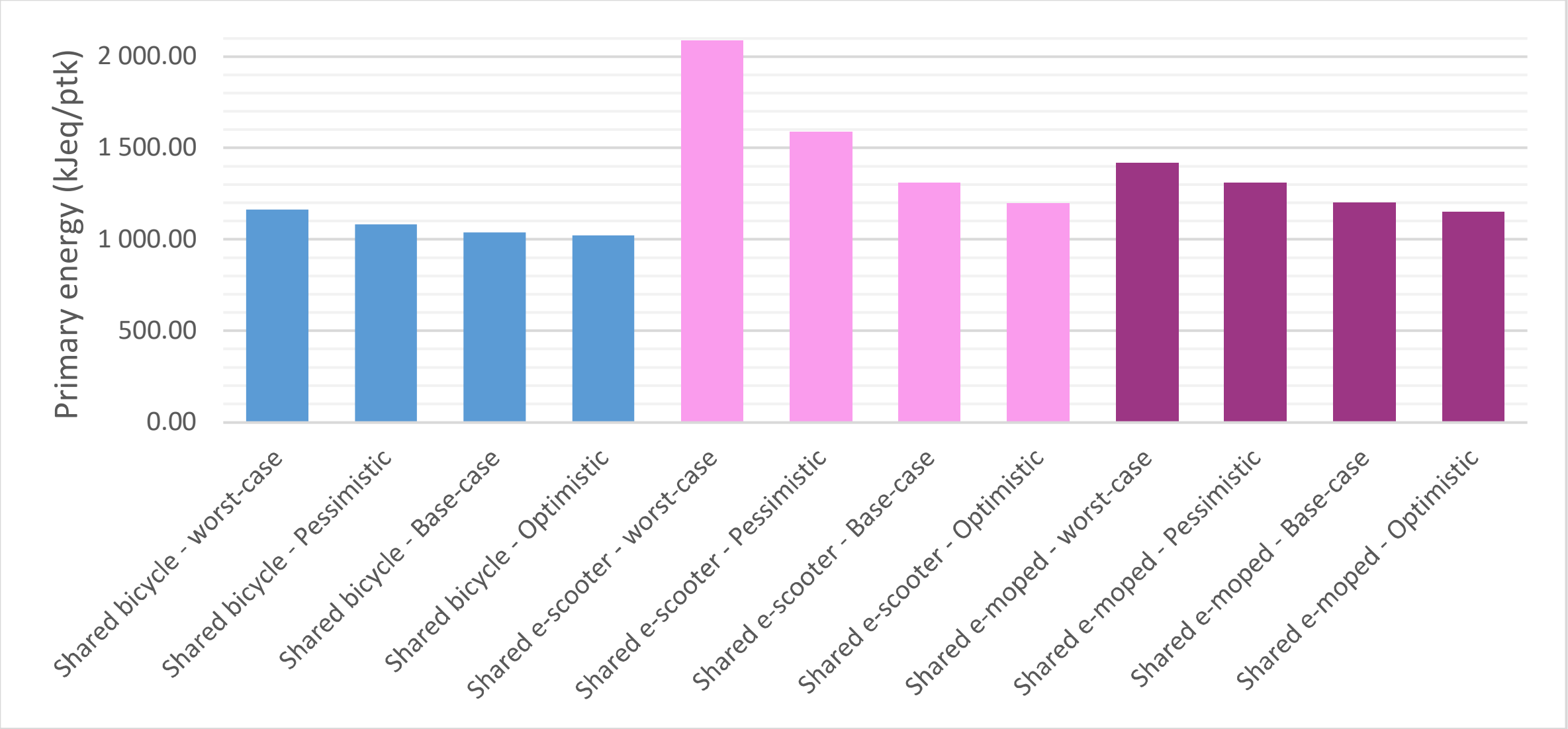

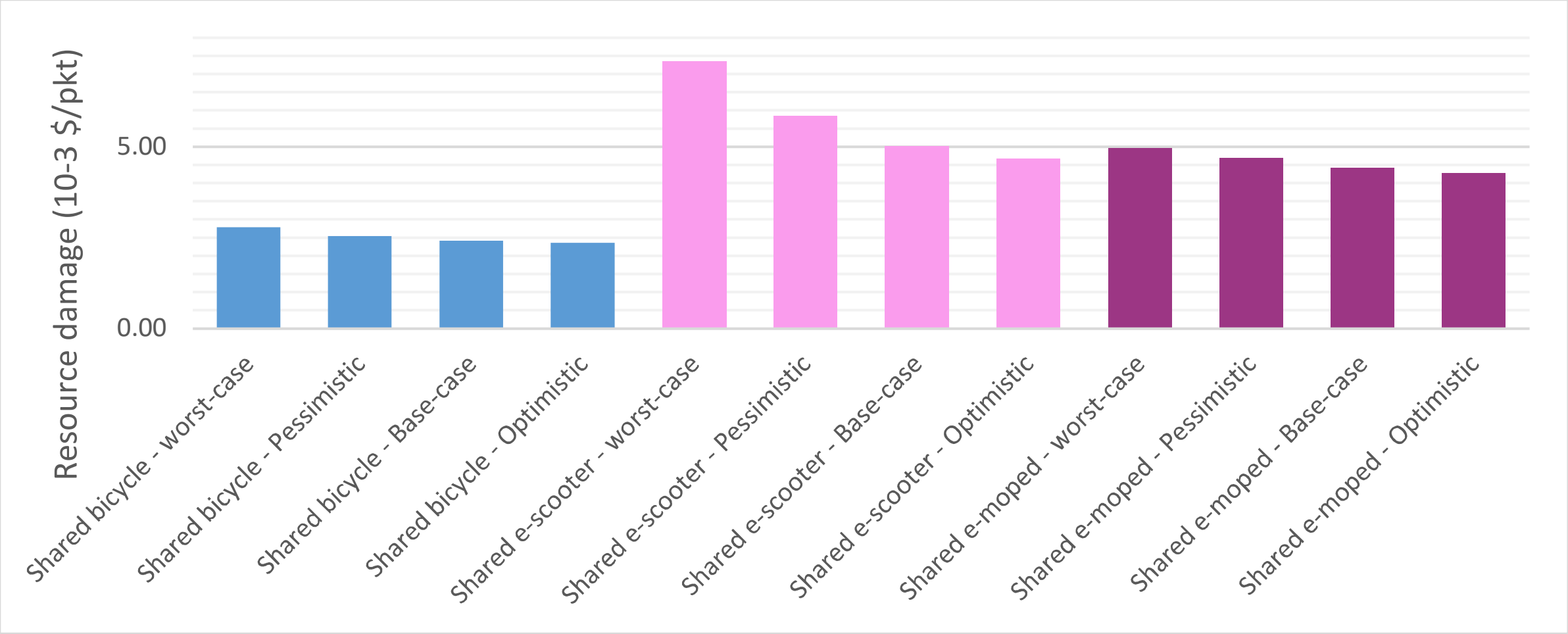

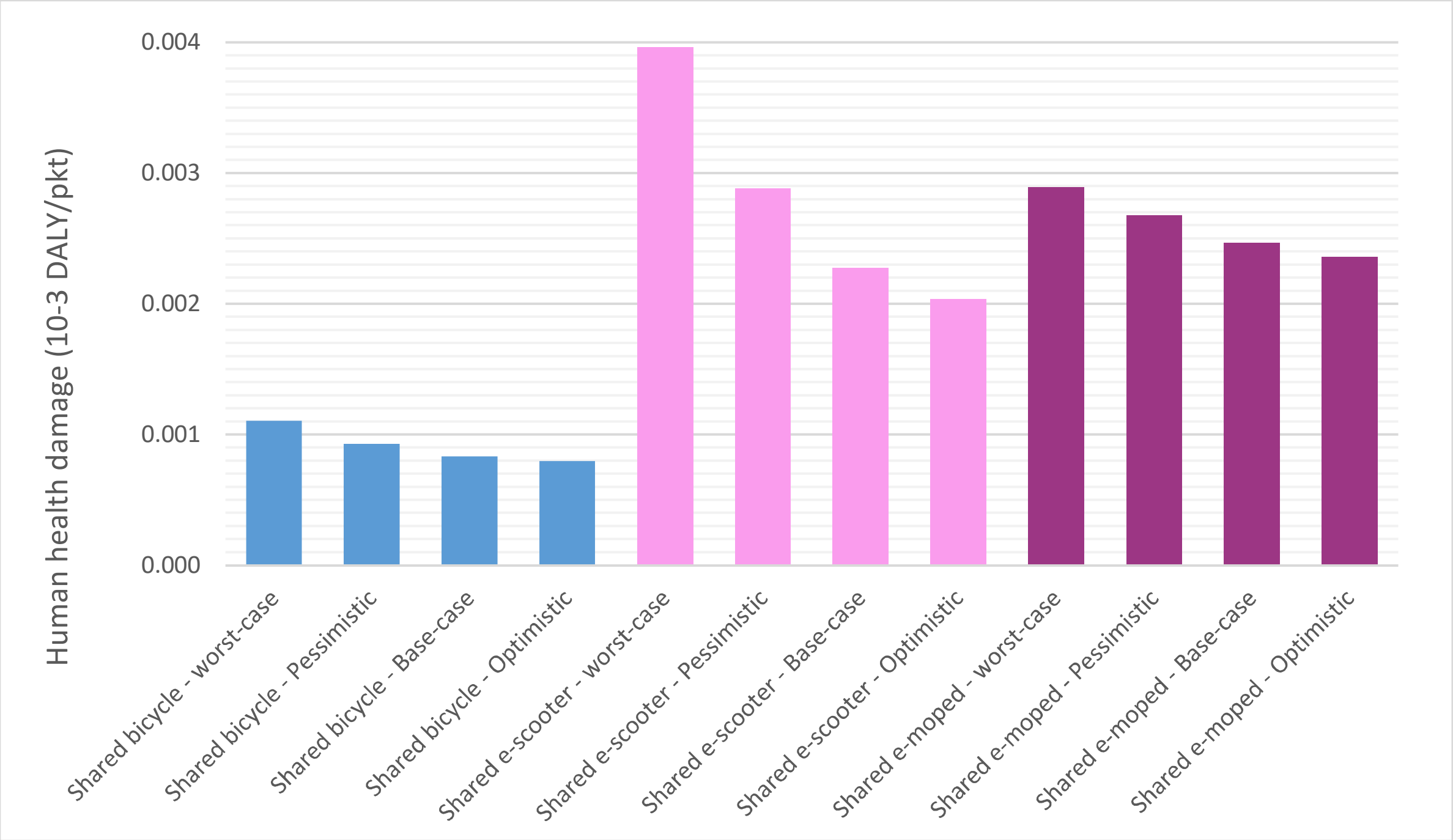

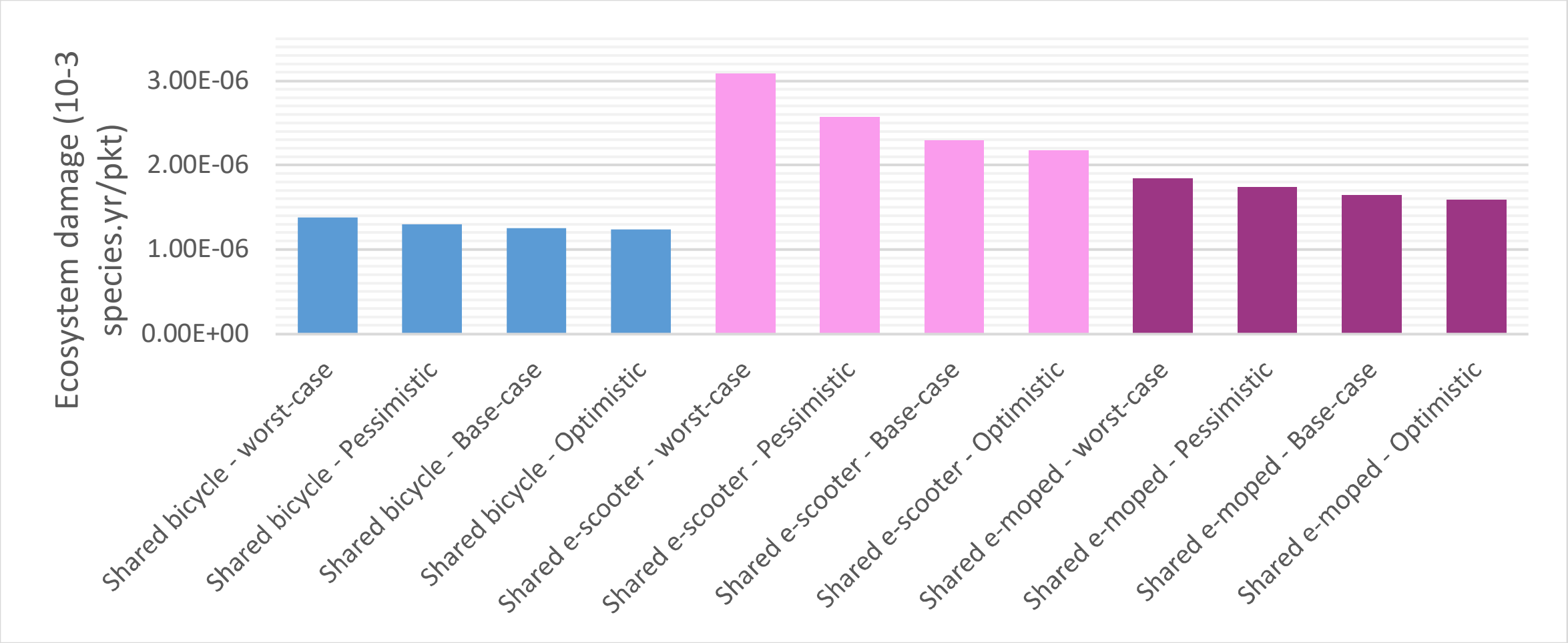

**Shipping scenarios**

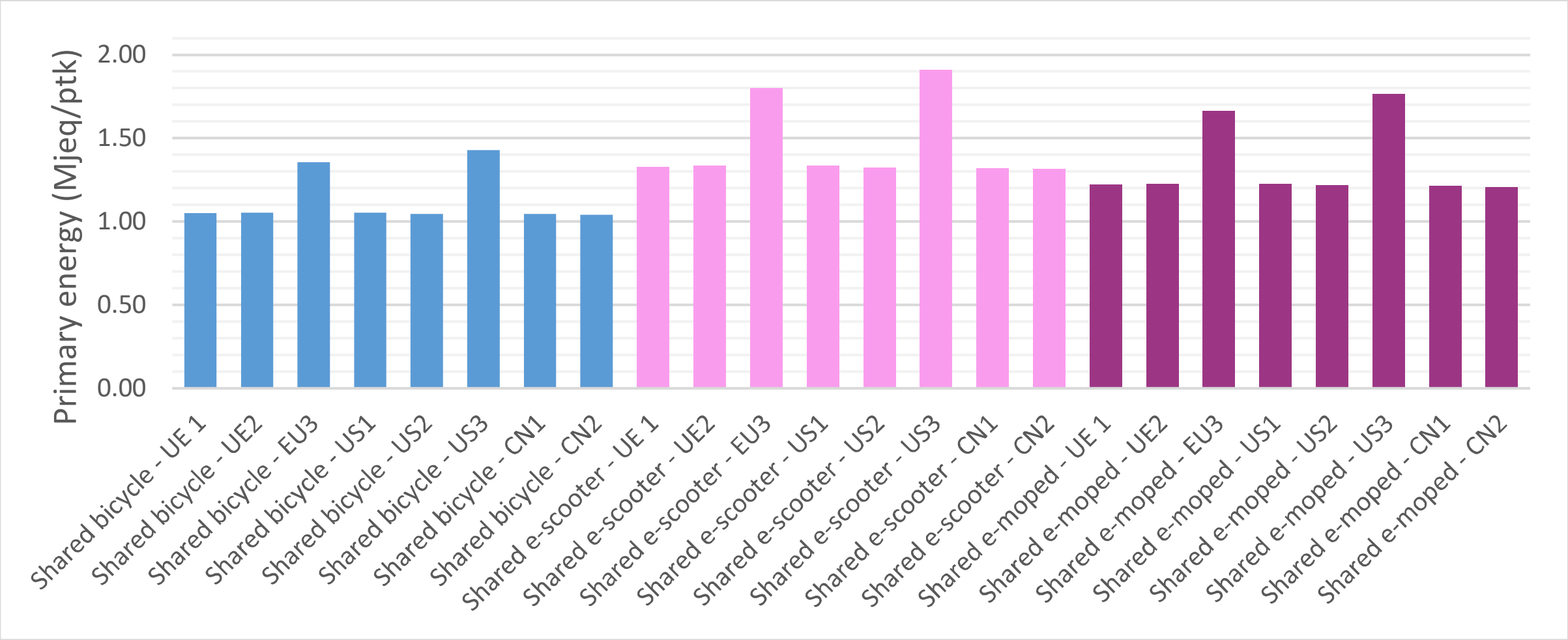

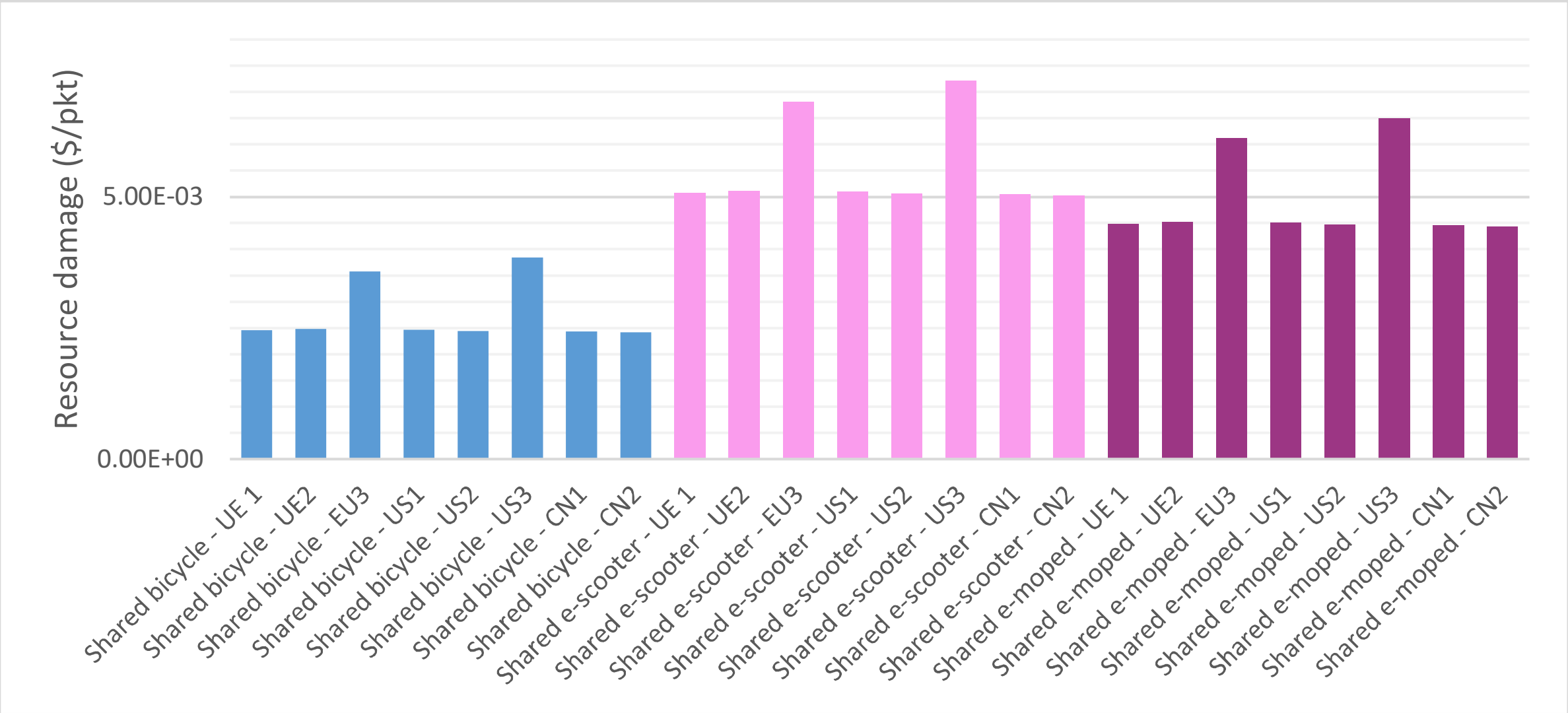

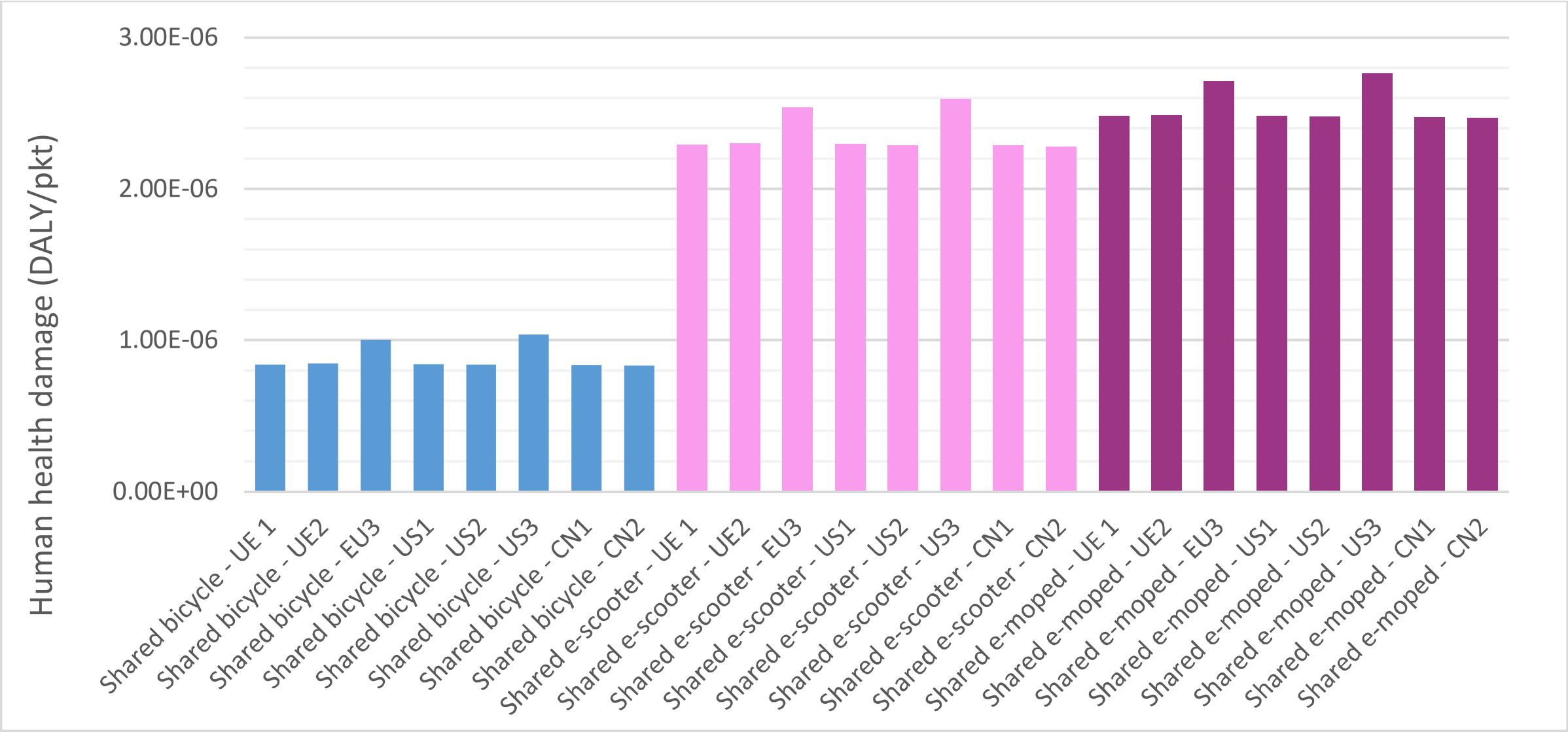

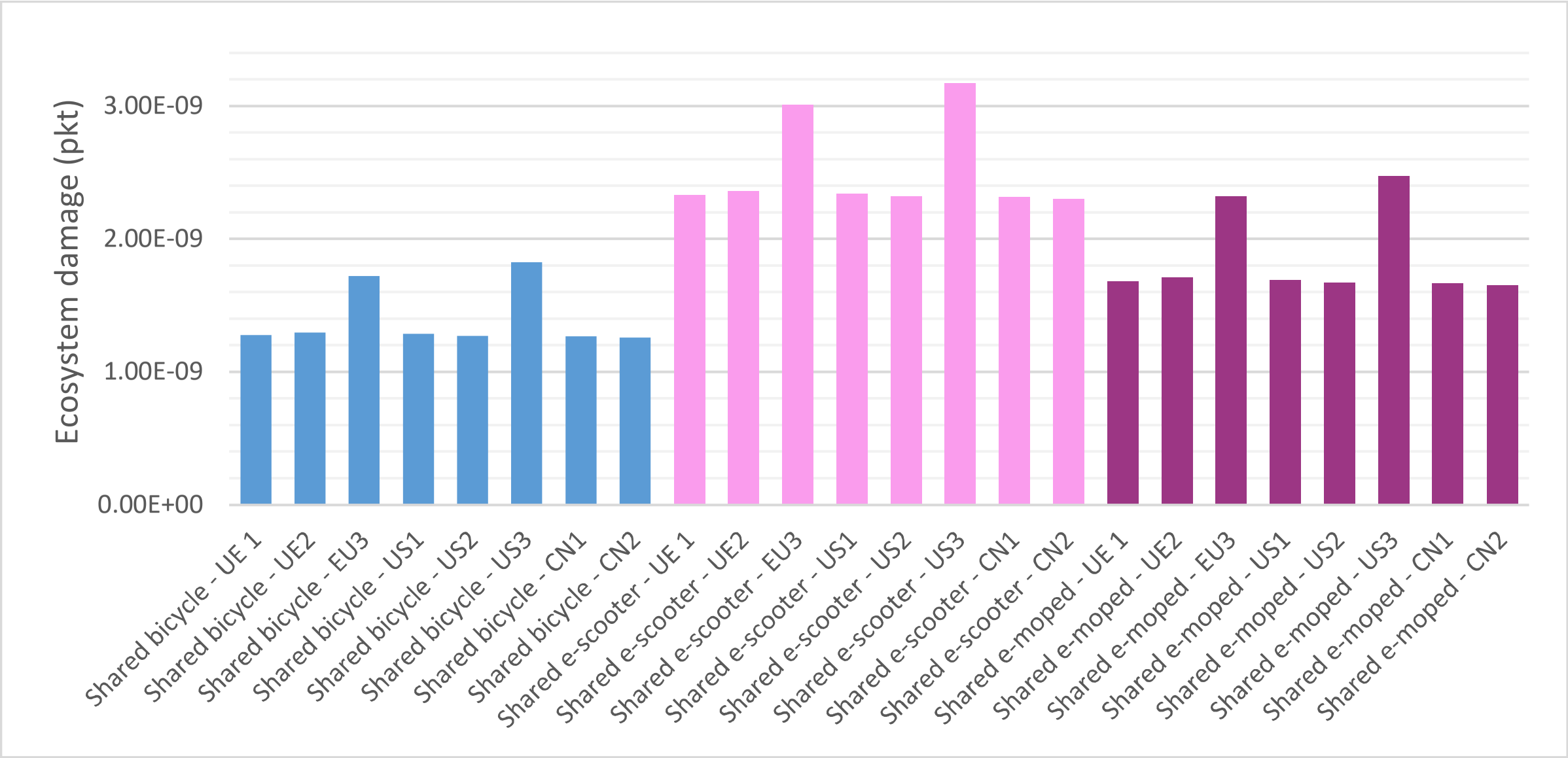

**Electricity mix**

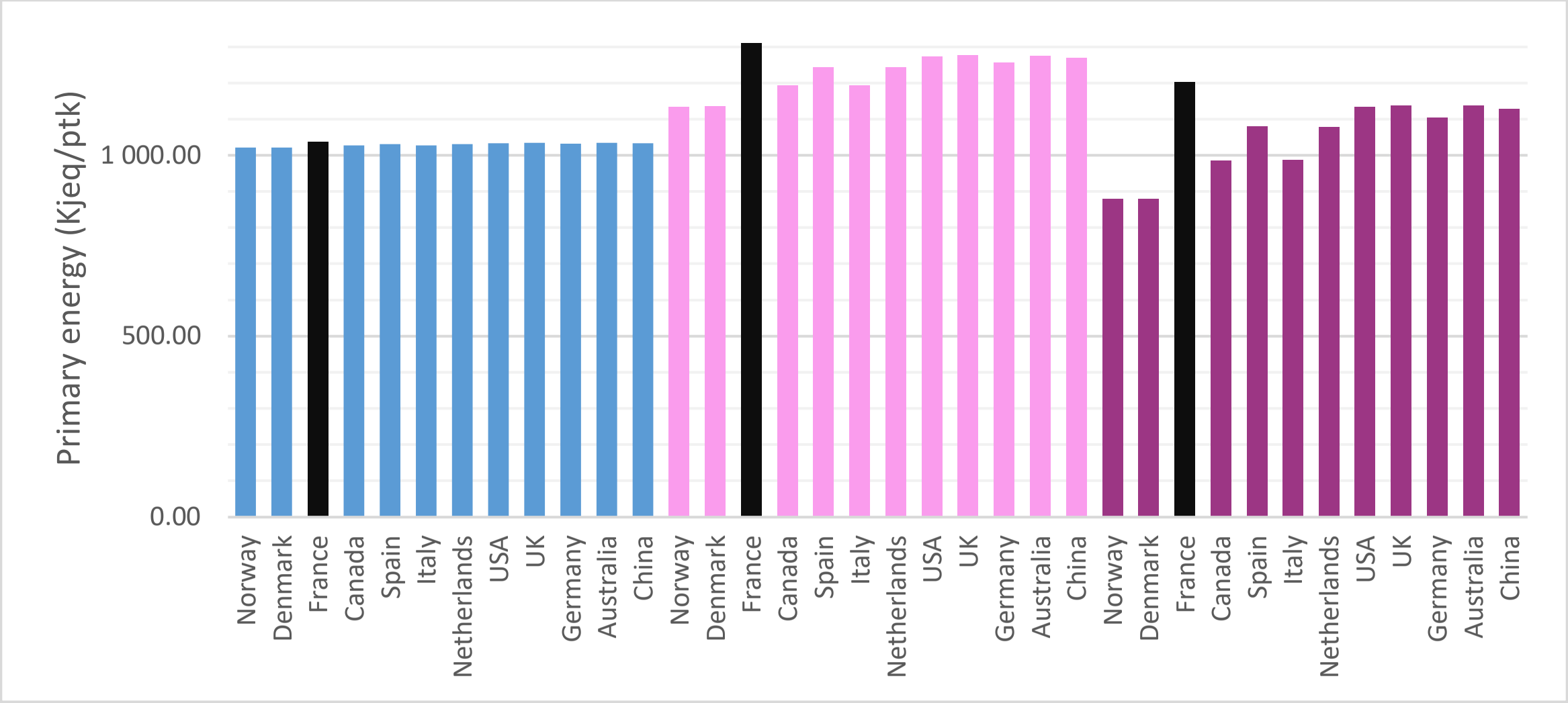

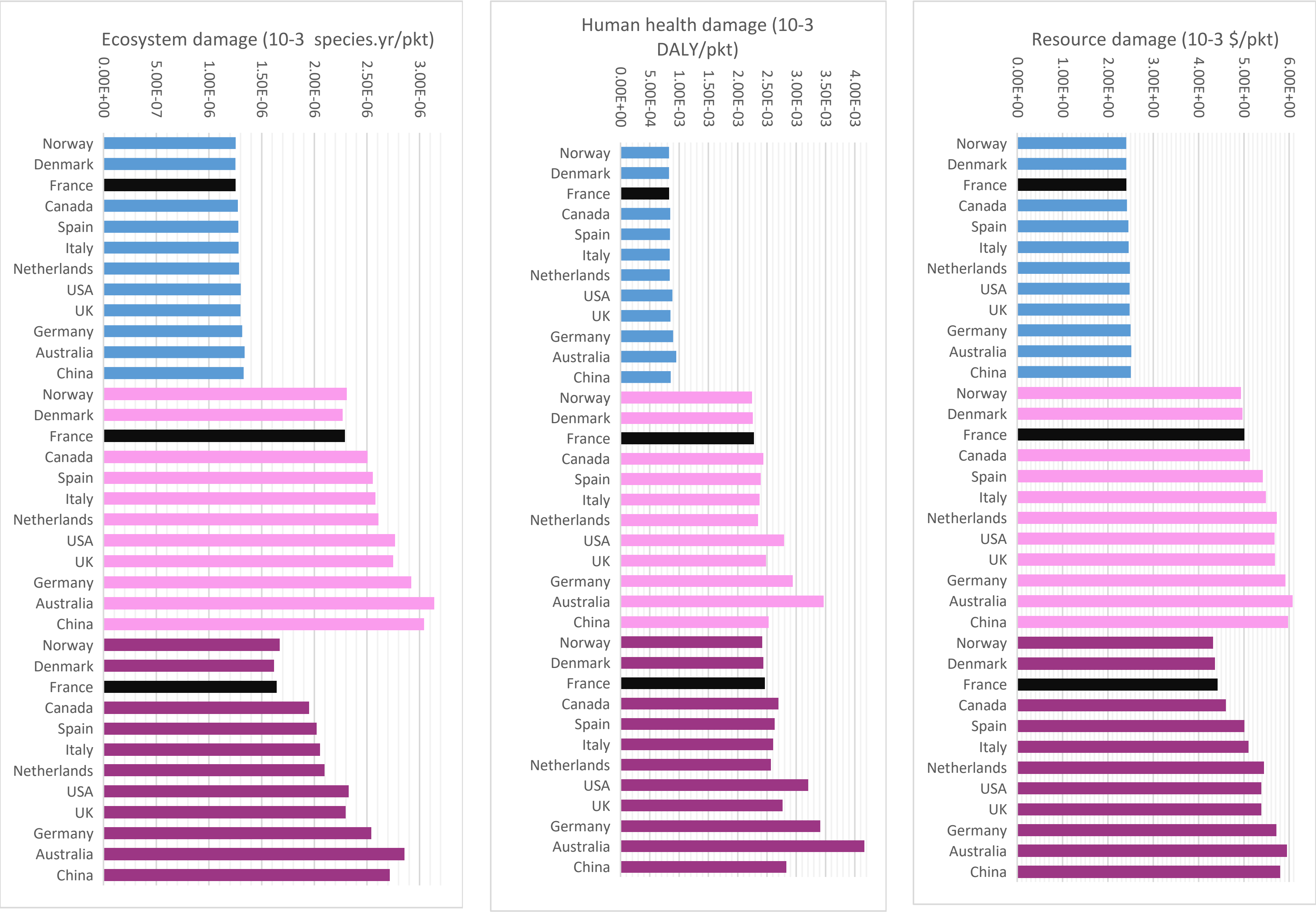

**Table 14 Contribution of the use and servicing stages to the total impact of the shared e-scooter per indicator in China**

| Shared e-scooter | CN use stage % | CN servicing stage % | total |
|---|---|---|---|
| **CC** | 22.47% | 18.64% | 41.11% |
| **EC** | 15.60% | 17.23% | 32.84% |
| **RD** | 11.34% | 17.18% | 28.52% |
| **HHD** | 8.03% | 22.75% | 30.78% |
| **ED** | 17.43% | 16.68% | 34.11% |

**Table 15 Contribution of the use and servicing stages to the total impact of the shared e-moped per indicator in China**

| e-moped | CN use stage % | CN servicing stage % | total |
|---|---|---|---|
| **CC** | 52.07% | 10.66% | 62.74% |
| **EC** | 34.60% | 9.36% | 43.96% |
| **RD** | 23.01% | 7.81% | 30.81% |
| **HHD** | 11.74% | 7.65% | 19.39% |
| **ED** | 41.12% | 9.60% | 50.72% |